\documentclass[12pt]{article}

\usepackage{latexsym}
\usepackage{amssymb,amsfonts,amsmath}
\usepackage{graphicx} 
\usepackage{indentfirst}
\usepackage{bbm}
\usepackage{amssymb}
\usepackage{verbatim}
\usepackage{amsmath, amsthm,amssymb}
\usepackage{mathrsfs}
\usepackage{hyperref}
\usepackage{amsfonts}
\usepackage{dsfont}
\usepackage{cite}
\usepackage{xcolor}
\usepackage[multiple]{footmisc}

\parskip=\medskipamount

\newcommand {\cC}{{\cal C}}
\newcommand {\cD}{{\cal D}}

\newcommand {\cF}{{\cal F}}

\newcommand {\cN}{{\cal N}}

\newcommand {\cP}{{\cal P}}
\newcommand {\cQ}{{\cal Q}}

\newcommand {\cS}{{\cal S}}

\newcommand {\cV}{{\cal V}}
\newcommand {\cW}{{\cal W}}

\def\a{\alpha}

\def\b{\beta}

\def\d{\delta}

\def\f{\phi}
\def\g{\gamma}

\def\j{\psi}

\def\l{\lambda}
\def\m{\mu}

\def\r{\rho}
\def\s{\sigma}
\def\t{\tau}

\def\z{\zeta}
\def\D{\Delta}
\def\F{\Phi}
\def\J{\Psi}
\def\L{\Lambda}
\def\O{\Omega}
\def\P{\Pi}

\def\U{\Upsilon}

\def\rd{{\rm d}}
\def\ri{{\rm i}}

\newcommand{\ad}{{\dot{\alpha}}}                           
\newcommand{\bd}{{\dot{\beta}}}                            
\newcommand{\ve}{\varepsilon}                            

\newcommand{\pa}{\partial}                           
\newcommand{\hf}{\frac12}

\newcommand{\vf}{\varphi}

\newcommand{\be}{\begin{equation}}
\newcommand{\ee}{\end{equation}}
\newcommand{\bea}{\begin{eqnarray}}
\newcommand{\eea}{\end{eqnarray}}
\newcommand{\non}{\nonumber}
\newcommand{\1}{{\underline{1}}}

\newcommand{\dsQ}{{\mathbb Q}}

\newcommand{\bm}[1]{\mbox{\boldmath$#1$}}

\def\double #1{#1{\hbox{\kern-2pt $#1$}}}

\newcommand{\mbS}{{\mathbb S}}

\newcommand{\mc}{\mathcal}
\newcommand{\mf}{\mathfrak}

\newcommand{\mb}{\mathbb}

\newif\ifdtup

\newcommand{\bsubeq}{\begin{subequations}}
\newcommand{\esubeq}{\end{subequations}}

\numberwithin{equation}{section}

\newcommand{\sSL}{\mathsf{SL}}

\newcommand{\sSO}{\mathsf{SO}}

\begin{document}

\begin{titlepage}
\begin{flushright}
July, 2021 \\
\end{flushright}
\vspace{5mm}

\begin{center}
{\Large \bf 
AdS (super)projectors in three dimensions and partial masslessness}
\\ 
\end{center}

\begin{center}

{\bf
Daniel Hutchings, Sergei M. Kuzenko and Michael Ponds} \\
\vspace{5mm}

\footnotesize{
{\it Department of Physics M013, The University of Western Australia\\
35 Stirling Highway, Crawley W.A. 6009, Australia}}  
~\\

\vspace{2mm}
~\\
\texttt{Email: daniel.hutchings@research.uwa.edu.au, sergei.kuzenko@uwa.edu.au, \\ michael.ponds@research.uwa.edu.au}
\vspace{2mm}

\end{center}

\begin{abstract}
\baselineskip=14pt

We derive the transverse projection operators for fields with arbitrary integer and half-integer spin on three-dimensional anti-de Sitter space, AdS$_3$.  
The projectors are constructed in terms of the quadratic Casimir operators of the 
isometry group $\mathsf{SO}(2,2)$ of  AdS$_3$.
Their poles are demonstrated 
to correspond to (partially) massless fields. 
As an application, we make use of the projectors to recast the conformal and topologically massive higher-spin actions in AdS$_3$ into a manifestly gauge-invariant and factorised form. 
We also propose operators which isolate the component of a field that is transverse and carries a definite helicity. Such fields correspond to irreducible representations of 
$\mathsf{SO}(2,2)$. 
Our results are then extended to the case of $\cN=1$ AdS$_3$ supersymmetry.
\end{abstract}
\vspace{5mm}

\vfill

\vfill
\end{titlepage}

\newpage
\renewcommand{\thefootnote}{\arabic{footnote}}
\setcounter{footnote}{0}

\tableofcontents{}
\vspace{1cm}
\bigskip\hrule

\allowdisplaybreaks
\newpage

\section{Introduction} 
The spin projection operators, or transverse and traceless (TT) spin-$s$ projectors, were first derived in four-dimensional (4d) Minkowski space $\mb{M}^4$ by Behrends and Fronsdal \cite{BF,F}. Given a symmetric tensor field on $\mb{M}^4$ that obeys the Klein-Gordon equation,  it decomposes into a sum of constrained fields describing irreducible representations of the Poincar\'e group with varying spin. The Behrends-Fronsdal projectors allow one to extract the component of this decomposition corresponding to the representation with the highest spin. Many applications for the TT projectors have been found within the landscape of high energy physics. For example, they played a crucial role in the original formulation of conformal higher-spin gauge actions \cite{FT}.

Since the work of \cite{BF,F}, the spin projection operators have been generalised to diverse dimensions and symmetry groups. In the case of $\mb{M}^d$, the TT projectors were first derived by Segal \cite{Se} (see also \cite{FMS, PoTs, B, IP}) in the bosonic case and later by Isaev and Podoinitsyn \cite{IP} for half-integer spins. In four dimensions, the projection operators in $\cN=1$ Minkowski superspace, $\mb{M}^{4|4}$, were introduced by Salam and Strathdee \cite{SalamS} in the case of a scalar superfield, and by Sokatchev \cite{Sokatchev}  for superfields of arbitrary rank. The superpojectors derived in \cite{Sokatchev} were formulated in terms of Casimir operators.  A few years later Rittenberg and Sokatchev \cite{RS} made use of a similar method to construct the  superprojectors in $\cN$-extended Minkowski superspace $\mb{M}^{4|4\cN}$ (see also \cite{Sokatchev2}).
An alternative powerful construction of the superprojectors
in $\mb{M}^{4|4\cN}$ was given in \cite{SG,GGRS}.\footnote{This approach has found numerous applications, e.g. the derivation of gauge-invariant  actions \cite{GS,GKP}.}
 Recently, the superprojectors in three-dimensional $\cN$-extended 
  Minkowski superspace,  ${\mathbb M}^{3|2\cN}$, 
 were derived in Ref. \cite{BHHK}, which built upon the earlier work of \cite{BLFKP}.

It is of interest to construct spin projection operators for fields on (anti-)de Sitter space, (A)dS. In particular, in order to describe  irreducible representations of the AdS$_d$ isometry algebra, $\mf{so}(d-1,2)$, fields on AdS$_d$ must satisfy certain differential constraints involving the  Lorentz-covariant derivative $\cD_a$ for AdS$_d$. Since both dS and AdS spaces have non-vanishing curvature, the construction of the TT projectors proves to be technically challenging. However, recent progress has been made in  \cite{KP4,BHKP}, where the (super)projectors in AdS$_4$ were derived. The next logical step is to derive the TT (super)projectors in AdS$_d$.  In this work we consider the case $d=3$, which serves as a starting point for this program.

This paper is organised as follows. In section \ref{secOSF}, we begin by reviewing on-shell fields in AdS$_3$ and the corresponding irreducible representations of $\mathfrak{so}(2,2)$ which they furnish. In section \ref{secBFProj}, we derive the spin projection operators for fields of arbitrary rank. More specifically, let us denote by $\cV_{(n)}$ the space of totally symmetric rank-$n$ spinor fields $\phi_{\a(n)}:=\phi_{\a_1\dots\a_n}=\phi_{(\a_1\dots\a_n)}$ on AdS$_3$. For any integer $n \geq 2$, we derive the rank-$n$ spin projection operator, $\Pi^{\perp}_{[n]}$, which is defined by its action on $\cV_{(n)}$ according to the rule:
\bea \label{Proj}
\Pi^{\perp}_{[n]}: \cV_{(n)} \longrightarrow \cV_{(n)}~, \qquad \f_{\a(n)} \longmapsto  \Pi^{\perp}_{[n]} \phi_{\a(n)}~  =:\phi^{\perp}_{\a(n)}~.
\eea
For fixed $n$, this operator is defined by the following properties:
\begin{enumerate}
	\item \textbf{Idempotence:} $\Pi^{\perp}_{[n]}$ is a projector in the sense that it squares to itself, 
	\bsubeq \label{SPO}
	\be \label{ProjProp}
	\Pi^{\perp}_{[n]}\Pi^{\perp}_{[n]}=\Pi^{\perp}_{[n]}~.
	\ee
	\item \textbf{Transversality:} $\Pi^{\perp}_{[n]}$ maps $\phi_{\a(n)}$ to a transverse field,
	\be \label{ProjTrans}
	 \cD^{\b(2)}\f^{\perp}_{\b(2)\a(n-2)} =0~.
	 \ee
	\item \textbf{Surjectivity:} 
	Every transverse field belongs to the image of $\Pi^{\perp}_{[n]}$, 
	\be \label{ProjU}
	\mc{D}^{\b(2)}\psi_{\b(2)\a(n-2)}=0~\quad\implies\quad\Pi^{\perp}_{[n]} \j_{\a(n)} = \j_{\a(n)}~.
	\ee
	\esubeq
	In other words, $\Pi^{\perp}_{[n]}$ acts as the identity operator on the space of transverse fields.
\end{enumerate}

Any operator satisfying all three of these properties may be considered to be an AdS$_3$ analogue of the Behrends-Fronsdal projector.\footnote{We refer to any operator satisfying properties \eqref{ProjProp}, \eqref{ProjTrans} and \eqref{ProjU} as a spin projection operator. In section \ref{secBFProj} we show that, under an additional assumption, such an operator is unique. In general, operators satisfying properties \eqref{ProjProp} and \eqref{ProjTrans} will be called transverse projectors. The latter are sometimes referred to as TT projectors, which is a slight abuse of terminology, since in vector notation the field $\f_{\a(n)}$ is already traceless.} However, the field $\phi^{\perp}_{\a(n)}$ will correspond to a reducible representation of $\mathfrak{so}(2,2)$.
In order to isolate the component describing an irreducible representation,  it is necessary to bisect the projectors according to $\Pi^{\perp}_{[n]}=\mb{P}_{[n]}^{(+)}+\mb{P}_{[n]}^{(-)}$. 
The operator $\mb{P}_{[n]}^{(\pm)}$ is a helicity projector since it satisfies the properties\footnote{Whilst $\mb{P}_{[n]}^{(\pm)}$ satisfies the properties \eqref{ProjProp} and \eqref{ProjTrans}, it does not satisfy \eqref{ProjU}. }
\eqref{ProjProp} and \eqref{ProjTrans} and selects the component of $\phi_{\a(n)}$ carrying the definite value $\pm\frac{n}{2}$ of helicity. They are constructed in section \ref{secHP}.
In section \ref{secLP} we make use of the orthogonal compliment of $\P^{\perp}_{[n]}$ to  decompose an unconstrained field $\f_{\a(n)}$ into a sum of transverse fields $\f^{\perp}_{\a(n-2j)}$ where $0 \leq j\leq \lfloor n/2 \rfloor$. We then provide an operator $\mbS^{\perp}_{\a(n-2j)}$ which extracts the field $\f^{\perp}_{\a(n-2j)}$ from this decomposition.

Making use of these projection operators, we derive
a number of interesting and non-trivial results. In particular, in section \ref{section2} we show that all information about (partially) massless fields is encoded in the poles of the transverse projectors. The novelty of our approach is that all projectors are derived in terms of the quadratic Casimir operators of $\mf{so}(2,2)$. This allows us to recast the AdS$_3$ higher-spin Cotton tensors and their corresponding conformal actions into a manifestly gauge-invariant and factorised form. Similar results are provided for new topologically massive (NTM) spin-$s$ gauge models, which are of order $2s$ in derivatives, where $s$ is a positive (half-)integer. In the case when $s$ is an integer, it is possible to construct NTM models of order $2s-1$. In $\mb{M}^3$ such models were recently proposed in \cite{DS}, here we extend them to AdS$_3$.
 The above results are detailed in section \ref{secCT}. Finally, in section \ref{secProjM} we study the flat limit of these results, and obtain new realisations for the spin projection operators, the helicity projectors and the conformal higher-spin actions in $\mb{M}^3$. 

 In section \ref{secSP}, we extend some of these results to the case of $\cN=1$ AdS$_3$ supersymmetry. Alongside concluding comments, new realisations of the Behrends-Fronsdal projectors in $\mb{M}^4$, expressed in terms of the Casimir operators of the 4d Poincar\'e algebra, are given in section \ref{section4}. The main body is accompanied by two technical appendices. Appendix \ref{appendixA} summarises our conventions and notation. We review the generating function formalism in Appendix \ref{appendixB}, which is a convenient framework used in deriving the non-supersymmetric results of section \ref{section2}.
 
 Our findings in this paper can be viewed as generalisations of the earlier results in
 AdS$_4$ \cite{KP4,BHKP} and AdS$_3$ \cite{KP1}, which in turn were based on the structure of (super)projectors in Minkowski (super)space \cite{BLFKP,BHHK}. Throughout this work we make use of the convention 
\begin{align}
U_{\a(n)} V_{\a(m)} = U_{(\a_1 . . .\a_n} V_{\a_{n+1} . . . \a_{n+m)}} ~.
\end{align}


\section{Transverse projectors in AdS$_3$} \label{section2}

The geometry of AdS$_3$ is described by the Lorentz covariant derivative,
\bea
\cD_{a}
=e_a{}^{m}\pa_m+\frac{1}{2}\omega_{a}{}^{bc}M_{bc}
=e_a{}^{m}\pa_m+\frac{1}{2}\omega_{a}{}^{\b\g}M_{\b\g}~,
\eea
 which satisfies the commutation relation
\be \label{ADSAlg}
[ \cD_a, \cD_b ] = -4 \cS^2 M_{ab} \quad \Longleftrightarrow \quad 
\ [ \cD_{\a \b}, \cD_{\g \d} ] = 4 \cS^2 \Big(\ve_{\g(\a}M_{\b)\d} + \ve_{\d(\a} M_{\b)\g}\Big)~.
\ee
Here $e_a{}^{m}$ is the inverse vielbein, $\omega_{a}{}^{bc}$ is the Lorentz connection and the parameter $\cS$ is related to the scalar curvature $R$ via  $R=-24\cS^2$. 
The Lorentz generators with vector ($M_{ab} =-M_{ba}$)
and spinor ($M_{\a\b} =M_{\b\a}$)  indices are defined in appendix
\ref{appendixA}.  In our subsequent analysis, we will make use of the quadratic Casimir operators of the AdS$_3$ isometry algebra $\mathfrak{so}(2,2) = \mathfrak{sl}(2,\mathbb{R}) \oplus \mathfrak{sl}(2,\mathbb{R})$, for which we choose (see, e.g., \cite{BPSS})
\bsubeq \label{QC}
\begin{alignat}{2} 
\mathcal{F}:&=\mathcal{D}^{\a\b}M_{\a\b}~,  &[\mathcal{F},\mathcal{D}_{\a\b}]=0~, \label{QCF}\\
	\mathcal{Q}:&= \Box -2\mathcal{S}^2M^{\a\b}M_{\a\b}~,  \qquad &[\mathcal{Q},\mathcal{D}_{\a\b}]=0~. \label{QCQ} 
\end{alignat}
\esubeq
Here $\Box:=\cD^a \cD_a = - \hf \cD^{\a\b}\cD_{\a\b}$ is the d'Alembert operator 
in AdS$_3$. The operators $\cF$ and $\cQ$  are related to each other as follows
\bea \label{FSQ}
\mathcal{F}^2 \f_{\a(n)} = n^2 \big [ \mathcal{Q} - (n-2)(n+2)\mathcal{S}^2 \big ]\f_{\a(n)} + n(n-1)\mathcal{D}_{\a(2)}\mathcal{D}^{\b(2)} \f_{\b(2)\a(n-2)}~,
\eea
for an arbitrary symmetric rank-$n$ spinor field $\f_{\a(n)}$.
The structure $\mathcal{D}_{\a(2)}\mathcal{D}^{\b(2)} \f_{\b(2)\a(n-2)}$ in \eqref{FSQ}
is not defined for the cases $n=0$ and $n=1$. However, it is multiplied by  $n(n-1)$ which  vanishes for these cases.


\subsection{On-shell fields}\label{secOSF}

In any irreducible representation of the AdS$_3$ isometry group $\sSO(2,2)$, the Casimir operators $\cF$ and $\cQ$  must be multiples of the identity operator. Therefore, in accordance with \eqref{FSQ}, one is led to consider on-shell fields of the type
\bsubeq \label{2.5}
\bea
\cD^{\b(2)}\f_{\b(2)\a(n-2)}&=&0~, \label{2.5a} \\
\big(\mc{F}-\m \big)\phi_{\a(n)}&=&0~, \label{2.5b}
\eea
\esubeq
for some real mass parameter $\m$.

Unitary representations of the Lie algebra $\mf{so}(2,2)$ may be realised in terms of the on-shell fields \eqref{2.5} for certain values of $\m$. As is well known (see, e.g., \cite{DKSS,BHRST} and references therein), the irreducible unitary representations of 
$\mf{so}(2,2)$ are denoted $D(E_0,s)$, where $E_0$ is the minimal energy (in units of $\mathcal{S}$), $s$ the helicity and $|s|$ is the spin. In this paper we are interested in only those representations carrying integer or half-integer spin with $|s|\geq 1$ and, consequently, the allowed values of $s$ are 
$s= \pm 1,\pm \frac{3}{2},\pm 2,\dots$~. In order for the representation $D(E_0, s)$ to be unitary, the inequality $E_0\geq |s|$, known as the unitarity bound, must be satisfied.

The representation $D(E_0,s)\equiv D(E_0,\s |s|)$, with $\s:=\pm 1$, may be realised on the space of symmetric rank-$n$ spinor fields $\phi_{\a(n)}$ satisfying the following differential constraints: 
\bsubeq \label{OC}
\bea
\cD^{\b(2)}\f_{\b(2)\a(n-2)}&=&0~, \label{OT} \\
\cD_{(\a_1}{}^{\b}\f_{\a_2 ... \a_n)\b} &=& \sigma\frac{\r}{n} \f_{\a(n)}~. \label{OM}
\eea
\esubeq
Here the integer $n\geq 2$ is related to $s$ via $ n=2|s|$. The real parameter 
$\rho \geq 0$, which carries mass dimension one, is called the pseudo-mass and is related to $E_0$ through
\begin{align}
E_0=1+\frac{\rho}{2n\mathcal{S}}~.
\end{align}
In terms of $\rho$ and $n$, the unitarity bound reads $\rho\geq n(n-2)\mc{S}$. With this in mind, we will label the representations using $\rho$ in place of $E_0$, and use the notation ${\mathfrak D}(\rho,\s \frac{n}{2})$.

The equations \eqref{OC} were introduced in 
\cite{BHRST}. In the flat-space limit, these equations 
reduce to those proposed in \cite{GKL,TV}.

The first-order equation  \eqref{OM} is equivalent 
to  \eqref{2.5b} with $\m = \s \r$.
Any field $\phi_{\a(n)}$ satisfying both constraints \eqref{OT} and \eqref{OM}, is an eigenvector of the Casimir operator $\mc{Q}$,
\begin{align}
\big(\mc{Q}-m^2\big)\phi_{\a(n)}=0~,\qquad m^2:=(\rho/n)^2+(n-2)(n+2)\mc{S}^2~. \label{OMS}
\end{align}

In place of \eqref{OT} and \eqref{OM}, one may instead consider tensor fields $\phi_{\a(n)}$ constrained by the equations \eqref{OT} and \eqref{OMS},
\bsubeq \label{OC2}
\bea
\cD^{\b(2)}\f_{\b(2)\a(n-2)}&=&0~, \label{OT2} \\
\big(\mc{Q}-m^2\big)\phi_{\a(n)}&=&0~. \label{OMS2}
\eea
\esubeq
In this case, the equation \eqref{FSQ} becomes
\begin{align}
\big(\mc{F}-\rho\big)\big(\mc{F}+\rho\big)\phi_{\a(n)}=0~.
\end{align}
It follows that such a $\phi_{\a(n)}$ furnishes the reducible representation 
\be \label{red}
{\mathfrak D}\Big (\rho,-\frac{n}{2} \Big )\oplus {\mathfrak D} \Big (\rho,\frac{n}{2} \Big )~.
\ee

It may be shown that when the pseudo-mass takes on any of the special values
\be
\r \equiv \r_{(t,n)}=  n(n-2t) \cS ~, \qquad 1 \leq t \leq \lfloor n/2 \rfloor~,
\label{PMvals}
\ee
then the representation ${\mathfrak D}(\rho,\s\frac{n}{2})$, with either sign for $\s$, shortens. At the field-theoretic level, this is manifested by the appearance of a depth-$t$ gauge symmetry
\begin{align}
\delta_{\z} \phi^{(t)}_{\a(n)}=\big(\mc{D}_{\a(2)}\big)^t\z_{\a(n-2t)}~, \label{PMGTs}
\end{align}
under which the system of equations \eqref{OC}, with $\rho$ given by \eqref{PMvals} and $\s$ arbitrary, is invariant.\footnote{This is true when the gauge parameter satisfies conditions analogous to \eqref{OC}, see \cite{KP1} for the details. } 
A field which satisfies the constraints \eqref{OT2} and \eqref{OMS}, and has pseudo-mass \eqref{PMvals}, will be said to be partially-massless with depth $t$ and denoted by $\phi^{(t)}_{\a(n)}$.\footnote{Partially massless fields have been studied in diverse dimensions for over 35 years, see e.g. \cite{DeserN1,Higuchi2,DW2,Zinoviev, Metsaev} for some of the earlier works. } 
For the field $\phi^{(t)}_{\a(n)}$ the second order equation \eqref{OMS} takes the form
\be
 \big ( \cQ - 
 \t_{(t,n)}\mc{S}^2
 \big )\f^{(t)}_{\a(n)}=0~,\qquad 
\t_{(t,n)}
= \big[2n(n-2t) +4(t-1)(t+1)\big]
~,\label{PMV}
\ee
where the parameters $\t_{(t,n)}$ are known as  the partially massless values. 
For $t>1$, the pseudo-mass  $\r_{(t,n)}$, eq. \eqref{PMvals},  violates the unitarity bound and hence the partially massless representations are non-unitary. 


\subsection{Spin projection operators} \label{secBFProj}

Given a tensor field $\f_{\a(n)}$ on AdS$_3$, the spin projection operator $\Pi^{\perp}_{[n]}$ with the defining properties \eqref{SPO}, selects the component $ \f^{\perp}_{\a(n)}$ of $\phi_{\a(n)}$ which is transverse. 
If, in addition, $\phi_{\a(n)}$ satisfies the second order mass-shell equation \eqref{OMS},
then $\Pi^{\perp}_{[n]}\phi_{\a(n)}$  furnishes the reducible representation ${\mathfrak D}(\rho,-\frac{n}{2})\oplus {\mathfrak D}(\rho,\frac{n}{2})$ of $\mf{so}(2,2)$.  

In this section we derive the spin projection operators $\Pi^{\perp}_{[n]}$.
 For this purpose it is convenient to make use of the generating function formalism, which is described in appendix \ref{appendixB}.
In this framework, the properties \eqref{ProjProp} and \eqref{ProjTrans} take the following form:
\be \label{PPGFF}
  \Pi^{\perp}_{[n]}\Pi^{\perp}_{[n]}\f_{(n)} = \Pi^{\perp}_{[n]}\f_{(n)}~,\qquad \mathcal{D}_{(-2)} \Pi^{\perp}_{[n]} {\f}_{(n)}=0~. \qquad 
\ee
It is necessary to separately analyse the cases with $n$ even and $n$ odd. 


\subsubsection{Bosonic case}

We will begin by studying the bosonic case, $n=2s$, for integer $s \geq 1$. Let us introduce the differential operator $\mathbb{T}_{[2s]}$
of order $2s$ in derivatives\footnote{When the upper bound in a product is less than the lower bound, we define the result to be unity. }
\be \label{BTO}
\mathbb{T}_{[2s]}=\sum_{j=0}^{s} 2^{2j}s\frac{(s+j-1)!}{(s-j)!}\prod_{t=1}^{j}\big (\mathcal{Q} - \t_{(s-t+1,2s)} \mathcal{S}^2 \big ) \mathcal{D}_{(2)}^{s-j}\mathcal{D}_{(-2)}^{s-j}~.
\ee
Here $\t_{(t,n)} $ denotes the partially massless values \eqref{PMV}. We refer the reader to appendix \ref{appendixB} for an explanation of the other notation. Given an arbitrary field $\f_{(2s)} \in \mc{V}_{(2s)}$, using \eqref{ID15} one may show that this operator maps it to a transverse field
\be
\mathcal{D}_{(-2)} \mathbb{T}_{[2s]} {\f}_{(2s)}=0~.
\ee
However, it is not a projector on $\mc{V}_{(2s)}$ since it does not square to itself,
\be
\mathbb{T}_{[2s]} \mathbb{T}_{[2s]} {\f}_{(2s)} = 2^{2s-1}(2s)! \prod_{t=1}^{s} \big (\mathcal{Q} - \t_{(t,2s)} \mathcal{S}^2 \big )\mathbb{T}_{[2s]} {\f}_{(2s)} ~.\label{Tsquared}
\ee  
To prove this identity, we observe that only the $j=s$ term of the sum  in \eqref{BTO} survives when $\mathbb{T}_{[2s]}$ acts on a transverse field such as $\mathbb{T}_{[2s]} {\f}_{(2s)} $. 

 To obtain a projector, we define the following dimensionless operator
\be \label{BP}
\widehat{\Pi}^{\perp}_{[2s]} := \Big [2^{2s-1}(2s)! \prod_{t=1}^{s} \big (\mathcal{Q} - \t_{(t,2s)} \mathcal{S}^2 \big )  \Big ]^{-1}\mathbb{T}_{[2s]} ~.
\ee
On $\mc{V}_{(2s)}$ it inherits its transversality from $\mathbb{T}_{[2s]}$, and is idempotent by virtue of \eqref{Tsquared}. In a fashion similar to the proof of \eqref{Tsquared}, it may also be shown that $\widehat{\Pi}^{\perp}_{[2s]}$ acts as the identity on the space of rank-$(2s)$ transverse fields.  Thus, $\widehat{\Pi}^{\perp}_{[2s]}$ satisfies the properties \eqref{SPO} and is hence the spin projection operator on $\mc{V}_{(2s)}$.
 Making the indices explicit, the latter reads
\bea \label{BosSpin}
\widehat{\Pi}^{\perp}_{[2s]}\f_{\a(2s)}  &=&\Big [ \prod_{t=1}^{s} \big (\mathcal{Q} - \t_{(t,2s)} \mathcal{S}^2 \big ) \Big ]^{-1} \sum_{j=0}^{s} 2^{2j-2s}\frac{2s}{s+j}\binom{s+j}{2j}\non \\
&&\times \prod_{t=1}^{j}\big (\mathcal{Q} - \t_{(s-t+1,2s)} \mathcal{S}^2 \big ) \mathcal{D}_{\a(2)}^{s-j}\big (\mathcal{D}^{\b(2)}\big )^{s-j}\f_{\a(2j)\b(2s-2j)}~. 
\eea

It is possible to construct a spin projection operator solely in terms of the two quadratic Casimir operators \eqref{QC}. To this end, we introduce the operator
 \begin{align} 
\P^{\perp}_{[2s]}= \frac{1}{2^{2s-1}(2s)!} \prod_{j=1}^{s}\frac{ \Big ( \cF^2 -4(j-1)^2 \big (\cQ-4j(j-2)\cS^2 \big ) \Big )}{ \big (\mathcal{Q} - \t_{(j,2s)} \mathcal{S}^2 \big ) }   
~. \label{SimpBosProj}
\end{align}
Let us show that \eqref{SimpBosProj} satisfies the three defining properties \eqref{SPO} on $\mc{V}_{(2s)}$. Given an arbitrary transverse field $\psi_{\a(2s)}$, $\mc{D}_{(-2)}\psi_{(2s)}=0$, using \eqref{FSQ} one may show that 
\bea
&&\prod_{j=1}^{s}\Big ( \cF^2 - 4  (j-1)^2 \big ( \cQ - 4j(j-2)\cS^2 \big ) \Big )  \psi_{(2s)} \non\\
&&= 2^{2s-1}(2s)!\prod_{j=1}^{s}\Big ( \cQ -\t_{(j,2s)} \cS^2 \Big) \psi_{(2s)} ~.\label{BosonicID1}
\eea
 It follows that ${\P}^{\perp}_{[2s]}$ acts as the identity on the space of transverse fields,
 \begin{align}
 \mc{D}_{(-2)}\psi_{(2s)}=0\quad \implies \quad \P^{\perp}_{[2s]}\psi_{(2s)}=\psi_{(2s)}~. \label{TTCPbos}
 \end{align}
 Next, the image of any unconstrained field $\f_{(2s)}$ under $\P^{\perp}_{[2s]}$ is transverse, which follows elegantly from \eqref{PTI}
\be
\cD_{(-2)} \P^{\perp}_{[2s]} \f_{(2s)} = \P^{\perp}_{[2s]} \cD_{(-2)} \f_{(2s)} \propto \mc{D}_{(2)}^{s}\mc{D}_{(-2)}^{s+1}\f_{(2s)}=0 ~. \label{TransBP}
\ee
 Finally, using \eqref{TTCPbos} and \eqref{TransBP} one can 
  show that $\P^{\perp}_{[2s]}$ squares to itself 
 \begin{align}
 \P^{\perp}_{[2s]}\P^{\perp}_{[2s]} \f_{(2s)} 
=\P^{\perp}_{[2s]} \f_{(2s)}~.  
 \end{align}
Thus $\P^{\perp}_{[2s]}$ satisfies \eqref{ProjProp}, \eqref{ProjTrans} and \eqref{ProjU} and can also be identified as a spin projector. 

Although it is not immediately apparent, the two projectors $\widehat{\P}^{\perp}_{[2s]}$ and $\P^{\perp}_{[2s]}$ actually coincide. Indeed, an operator satisfying the three properties \eqref{SPO}, and which commutes with $\cD_a$, must be unique. Let us explain why this is so. Take an arbitrary $\phi_{(2s)}$ and act on it first with $\widehat{\P}^{\perp}_{[2s]}$ and then with $\P^{\perp}_{[2s]}$. Since $\widehat{\P}^{\perp}_{[2s]}\phi_{(2s)}$ is transverse, and $\P^{\perp}_{[2s]}$ acts as the identity on this space, we have
\begin{align}
\P^{\perp}_{[2s]}\widehat{\P}^{\perp}_{[2s]}\f_{(2s)} = \widehat{\P}^{\perp}_{[2s]}\f_{(2s)}~. \label{Subtract1}
\end{align}
Next, we perform the same operation but in the opposite order, 
\begin{align}
\widehat{\P}^{\perp}_{[2s]}\P^{\perp}_{[2s]}\f_{(2s)} = \P^{\perp}_{[2s]}\f_{(2s)}~, \label{Subtract2}
\end{align}
and subtract \eqref{Subtract1} from \eqref{Subtract2}. Using the fact that $\P^{\perp}_{[2s]}$ is composed solely from Casimir operators, and hence commutes with $\widehat{\P}^{\perp}_{[2s]}$, it follows that on $\mc{V}_{(2s)}$ the two are equal to one another,
\be \label{EBP}
\widehat{\P}^{\perp}_{[2s]}\f_{(2s)}=\P^{\perp}_{[2s]}\f_{(2s)} ~.
\ee

So far our analysis of the spin projection operators $\widehat{\Pi}_{[2s]}^{\perp}$ and $\Pi_{[2s]}^{\perp}$ has been restricted to the linear space $\mc{V}_{(2s)}$. However, for fixed $s$,
 the operator $\Pi_{[2s]}^{\perp}$ given by  eq. \eqref{SimpBosProj}  is also defined 
 to act on the linear spaces $\mc{V}_{(2s')}$ with $s' < s$. 
In fact, making use of \eqref{FSQ} and \eqref{PTI}, it is possible to show that the following holds true 
\be \label{BosProjProp}
\P^{\perp}_{[2s]} \f_{(2s')}=0~, \qquad 1 \leq s' \leq s-1 ~.
\ee 
This important identity states that $\Pi^{\perp}_{[2s]}$ annihilates any lower-rank field $\f_{\a(2s')}\in \mc{V}_{(2s')}$. It should be mentioned that $\P^{\perp}_{[2s]} $ does not annihilate lower-rank fermionic fields
 $\f_{\a(2s'+1)}$.
When acting on $\cV_{(2s')}$, the two operators $\widehat{\Pi}_{[2s]}^{\perp}$ and $\Pi_{[2s]}^{\perp}$ are no longer equal to each other, and in particular $\widehat{\Pi}_{[2s]}^{\perp}\phi_{(2s')}\neq 0 $. It is for this reason that we will continue to use different notation for the two operators. 

It follows from \eqref{SimpBosProj} that the poles of  $\P^{\perp}_{[2s]} $ correspond to the partially massless values $\t_{(j,2s)}$ defined by \eqref{PMV}.


\subsubsection{Fermionic case}

We now turn our attention to the fermionic case, $n=2s+1$,  for integers $s\geq 1$. Let us introduce the differential operator $\mathbb{T}_{[2s+1]}$ of order $2s$ in derivatives
\be \label{FTO}
\mathbb{T}_{[2s+1]}=\sum_{j=0}^{s} 2^{2j}\frac{(s+j)!}{(s-j)!}\prod_{t=1}^{j}\big (\mathcal{Q} - \t_{(s-t+1,2s+1)} \mathcal{S}^2 \big ) \mathcal{D}_{(2)}^{s-j}\mathcal{D}_{(-2)}^{s-j}~.
\ee
Here $\t_{(t,n)}$ are the partially massless values \eqref{PMV}. The operator $\mathbb{T}_{[2s+1]}$ maps $\f_{(2s+1)}$ to a transverse field
\be
\mathcal{D}_{(-2)} \mathbb{T}_{[2s+1]} {\f}_{(2s+1)}=0~.
\ee
However, this operator does not square to itself on $\cV_{(2s+1)}$
\be \label{TFsquared}
\mathbb{T}_{[2s+1]}\mathbb{T}_{[2s+1]}\f_{(2s+1)} = 2^{2s}(2s)! \prod_{t=1}^{s} \big (\mathcal{Q} - \t_{(t,2s+1)} \mathcal{S}^2 \big ) \mathbb{T}_{[2s+1]}\f_{(2s+1)} ~.
\ee
As a result, one can immediately define the dimensionless operator
\be \label{FP}
\widehat{\P}^{\perp}_{[2s+1]}  := \Big [2^{2s}(2s)! \prod_{t=1}^{s} \big (\mathcal{Q} - \t_{(t,2s+1)} \mathcal{S}^2 \big ) \Big ]^{-1} ~\mathbb{T}_{[2s+1]} ~,
\ee
which is a transverse projector by construction. Following a derivation similar to that of \eqref{TFsquared}, it can be shown that the operator $\widehat{\P}^{\perp}_{[2s+1]}$ acts like the identity on the space of transverse fields. Hence, the operator $\widehat{\P}^{\perp}_{[2s+1]}$ satisfies properties \eqref{SPO}, and is thus a spin projection operator on $\cV_{(2s+1)}$. Converting \eqref{FP} to spinor notation yields
\bea \label{FermSpin}
\widehat{\P}^{\perp}_{[2s+1]} \f_{\a(2s+1)}&=&\Big [\prod_{t=1}^{s} \big (\mathcal{Q} - \t_{(t,2s+1)} \mathcal{S}^2 \big ) \Big ]^{-1} \sum_{j=0}^{s} 2^{2j-2s}\frac{2s+1}{2j+1}\binom{s+j}{2j}~ \non\\
&&\times \prod_{t=1}^{j}\big (\mathcal{Q} - \t_{(s-t+1,2s+1)} \mathcal{S}^2 \big )   \mathcal{D}_{\a(2)}^{s-j}\big (\mathcal{D}^{\b(2)}\big )^{s-j}\f_{\a(2j+1)\b(2s-2j)}~.
\eea

As in the bosonic case, one can construct a fermionic projector purely in terms of the quadratic Casimir operators \eqref{QC}. Let us introduce the operator 
\bea 
{\P}^{\perp}_{[2s+1]} 
= \frac{1}{2^{2s}(2s)!}  \prod_{j=1}^{s} \frac{\Big ( \cF^2 -(2j-1)^2 \big (\cQ-(2j-3)(2j+1)\cS^2 \big ) \Big )}{\big (\mathcal{Q} - \t_{(j,2s+1)} \mathcal{S}^2 \big ) }
~. \label{SimpFermProj}
\eea
We wish to show that \eqref{SimpFermProj} indeed satisfies the properties \eqref{SPO} on $\cV_{(2s+1)}$. Given an arbitrary transverse field $\j_{(2s+1)} $, 
using \eqref{FSQ} one can derive the identity
\bea \label{RFP}
&&\prod_{j=1}^{s}\Big ( \cF^2 - \big (2j-1 \big )^2 \big ( \cQ - (2j-3)(2j+1)\cS^2 \big ) \Big ) \j_{(2s+1)} \\
&&= 2^{2s}(2s)!\prod_{j=1}^{s}\Big ( \cQ -\t_{(j,2s+1)} \cS^2 \Big )  \j_{(2s+1)} ~. \non 
\eea
It follows that ${\P}^{\perp}_{[2s+1]}$ acts like the identity on the space of transverse fields
\be \label{FPI}
\cD_{(-2)}\j_{(2s+1)} = 0 \quad \Longrightarrow \quad   {\P}^{\perp}_{[2s+1]}\j_{(2s+1)}  = \j_{(2s+1)} ~.
\ee
By making use of \eqref{PTI}, one can show that the operator $ {\P}^{\perp}_{[2s+1]}$ maps $\f_{(2s+1)}$ to a transverse field
\be
\cD_{(-2)} {\P}^{\perp}_{[2s+1]} \f_{(2s+1)} = {\P}^{\perp}_{[2s+1]} \cD_{(-2)} \f_{(2s+1)} \propto \mc{D}_{(2)}^{s}\mc{D}_{(-2)}^{s+1}\f_{(2s+1)}=0 ~.\label{TransFP}
\ee
Finally, using \eqref{FPI} in conjunction with \eqref{TransFP}, one can show that $ {\P}^{\perp}_{[2s+1]}$  is idempotent
\begin{subequations}
	\bea
	{\P}^{\perp}_{[2s+1]}{\P}^{\perp}_{[2s+1]} \f_{(2s+1)} ={\P}^{\perp}_{[2s+1]} \f_{(2s+1)} ~.
	\eea
\end{subequations}
Hence, $ {\P}^{\perp}_{[2s+1]}$ satisfies \eqref{SPO}, and can thus be classified as a spin projector on AdS$_3$.

In a similar fashion to the bosonic case, it may be shown that $\widehat{\P}^{\perp}_{[2s+1]}$ and ${\P}^{\perp}_{[2s+1]}$ are equivalent on $\cV_{(2s+1)}$,
\be  \label{EFP}
\widehat{\Pi}^{\perp}_{[2s+1]}\f_{(2s+1)} ={\Pi}^{\perp}_{[2s+1]}\f_{(2s+1)}  ~.
\ee
Stepping away from $\cV_{(2s+1)}$, one can show that for fixed $s$, the projector $\Pi_{[2s+1]}^{\perp}$ annihilates any lower-rank field $\f_{(2s'+1)}\in \mc{V}_{(2s'+1)}$
\be \label{FermProjProp}
\P^{\perp}_{[2s+1]} \f_{(2s'+1)}=0~, \qquad 1 \leq s' \leq s-1 ~.
\ee 
The two operators $\widehat{\Pi}_{[2s+1]}^{\perp}$ and $\Pi_{[2s+1]}^{\perp}$ are not equivalent on $\mc{V}_{(2s'+1)}$. We remark that $\P^{\perp}_{[2s+1]} $ does not annihilate lower-rank bosonic fields $\f_{\a(2s'+2)}$.

It follows from \eqref{SimpFermProj} that the poles of  $\P^{\perp}_{[2s+1]} $ correspond to the partially massless values $\t_{(j,2s+1)}$ defined by \eqref{PMV}.

An important property of the projectors \eqref{SimpBosProj} and  
\eqref{SimpFermProj} is that they are symmetric operators, that is 
\bea
\int\text{d}^3x\, e \, \j^{\a(n)} \P^{\perp}_{[n]} \f_{\a(n)} 
= \int\text{d}^3x\, e \, \f^{\a(n)} \P^{\perp}_{[n]} \j_{\a(n)} ~, \qquad 
e^{-1}:= \text{det}(e_a{}^{m})~,
\eea
for arbitrary well-behaved fields $\j_{\a(n)} $ and $\f_{\a(n)} $.


\subsection{Helicity projectors} \label{secHP}

As previously mentioned,  given a rank-$n$ field $\phi_{\a(n)}$ satisfying the mass-shell equation \eqref{OMS}, its projection $\Pi_{[n]}^{\perp}\phi_{\a(n)}$ furnishes the reducible representation ${\mathfrak D}(\rho,-\frac{n}{2})\oplus {\mathfrak D}(\rho,\frac{n}{2})$. In particular, representations with both signs of helicity $\pm\frac{n}{2}$ appear in this decomposition. 
 
 In order to isolate the component of $\phi_{\a(n)}$ describing an irreducible representation of $\mf{so}(2,2)$, it is necessary to split the spin projection operators $\Pi_{[n]}^{\perp}$ according to
 \begin{align}
 \Pi_{[n]}^{\perp}=\mathbb{P}^{(+)}_{[n]}+\mathbb{P}^{(-)}_{[n]}~.\label{splitting}
 \end{align}    
Each of the helicity projectors $\mathbb{P}^{(\pm)}_{[n]}$ should satisfy \eqref{ProjProp} and \eqref{ProjTrans}. In addition, they should project out the component of $\phi_{\a(n)}$ carrying a single value of helicity. The last two requirements are equivalent to the equations
\begin{subequations}
\begin{align}
\mc{D}^{\b(2)}\f^{(\pm)}_{\b(2)\a(n-2)}&=0~,\\
\big(\mc{F}\mp\rho\big)\f^{(\pm)}_{\a(n)}&=0~,\label{HPprop2}
\end{align}
\end{subequations}
where we have denoted $\f^{(\pm)}_{\a(n)}:=\mathbb{P}^{(\pm)}_{[n]} \f_{\a(n)}$. It follows that $\f^{(\pm)}_{\a(n)}$ furnishes the irreducible representation ${\mathfrak D}(\rho,\pm \frac{n}{2})$.

It is not difficult to show that the following operators satisfy these requirements
\be \label{helicityproj}
\mathbb{P}^{(\pm)}_{[n]} :=\hf \bigg (\mathds{1} \pm \frac{\mathcal{F}}{n\sqrt{\mathcal{Q}-(n+2)(n-2)\mathcal{S}^2}} \bigg ) {\Pi}^{\perp}_{[n]}~.
\ee 
Here ${\Pi}^{\perp}_{[n]}$ are the spin projectors written in terms of Casimir operators, and are given by \eqref{SimpBosProj} and \eqref{SimpFermProj} in the bosonic and fermionic cases respectively. 
 Of course, on $\cV_{(n)}$, one could instead choose to represent the latter in their alternate form \eqref{BP} and \eqref{FP}.

Using the defining features of ${\Pi}^{\perp}_{[n]}$, it can be shown that the operators $\mathbb{P}^{(+)}_{[n]} $ and $\mathbb{P}^{(-)}_{[n]} $ are orthogonal projectors when restricted to $\mc{V}_{(n)}$:
\be \label{OHP}
\mathbb{P}^{(\pm)}_{[n]} \mathbb{P}^{(\pm)}_{[n]}  = \mathbb{P}^{(\pm)}_{[n]} ~, \qquad \mathbb{P}^{(\pm)}_{[n]} \mathbb{P}^{(\mp)}_{[n]}  = 0~.
\ee
It is also clear that \eqref{helicityproj} projects onto the transverse subspace of $\mc{V}_{(n)}$-- it inherits this property from ${\Pi}_{[n]}$. Moreover, the off-shell field $\f^{(\pm)}_{\a(n)}$ satisfies the constraint
\be \label{helcityprojprop}
\Big(\mathcal{F} \mp n \sqrt{\mathcal{Q} - (n-2)(n+2)\mathcal{S}^2} \Big)\f^{(\pm)}_{\a(n)} =0 ~.
\ee
If $\f^{(\pm)}_{\a(n)}$ is on the mass-shell, eq. \eqref{OMS}, then \eqref{helcityprojprop} reduces to \eqref{HPprop2}.


\subsection{Longitudinal projectors and lower-spin extractors} \label{secLP}

In this section we study the operator $\Pi^{\parallel}_{[n]}$ which is the compliment of $\P^{\perp}_{[n]}$,
\be \label{LongProj}
\Pi^{\parallel}_{[n]}  :=  \mathds{1} - \P^{\perp}_{[n]} ~.
\ee
By construction, the two operators $\Pi^{\perp}_{[n]}$ and $\Pi^{\parallel}_{[n]}$ resolve the identity, $\mathds{1} = \Pi^{\parallel}_{[n]} + \Pi^{\perp}_{[n]}$, and form an orthogonal set of projectors
\begin{subequations} \label{OrthoProjProp}
	\begin{align}
	\Pi^{\perp}_{[n]}\Pi^{\perp}_{[n]} &= \Pi^{\perp}_{[n]}~,\qquad \Pi^{\parallel}_{[n]}\Pi^{\parallel}_{[n]} = \Pi^{\parallel}_{[n]}~, \label{DualProps}\\
	\Pi^{\parallel}_{[n]} \Pi^{\perp}_{[n]} &=0~, \qquad  ~~~\phantom{.} \Pi^{\perp}_{[n]} \Pi^{\parallel}_{[n]}=0 ~.
	\end{align}
\end{subequations}
Moreover, it can be shown that $\Pi^{\parallel}_{[n]}$ projects a field $\f_{\a(n)}$ onto its longitudinal component. A rank-$n$ field $\psi_{\a(n)}$ is said to be longitudinal if there exists a rank-$(n-2)$ field $\psi_{\a(n-2)}$ such that $\psi_{\a(n)}$ may be expressed as $\psi_{\a(n)} = \mathcal{D}_{\a(2)}\psi_{\a(n-2)}$. Such fields are also sometimes referred to as being pure gauge.
Therefore, we find that 
\begin{align}
\f^{\parallel}_{\a(n)}:=\Pi^{\parallel}_{[n]} \f_{\a(n)}=\mc{D}_{\a(2)}\f_{\a(n-2)}~,
\end{align}
for some unconstrained field $\phi_{\a(n-2)}$. For $\phi_{\a(n)}$ off-shell, $\f_{\a(n-2)}$ will be non-local in general. For example, in the case of a vector field $\phi_a$, we have $\phi^{\parallel}_a=\mc{D}_a\phi$ where $\phi=\frac{1}{\mc{Q}}\mc{D}^a\phi_a$.

Using the fact that $\Pi^{\perp}_{[n]}$ and $\Pi^{\parallel}_{[n]}$ resolve the identity, one can decompose an arbitrary field $\f_{\a(n)}$ as follows
\be
\f_{\a(n)} = \f^{\perp}_{\a(n)} + \cD_{\a(2)}\f_{\a(n-2)}~.
\label{2.51}
\ee
Here $\f^{\perp}_{\a(n)}$ is transverse and $\f_{\a(n-2)}$ is unconstrained. Repeating this process iteratively, we obtain the following decomposition
\bea\label{Decomp}
\f_{\a(n)} &=& \sum_{j=0}^{\lfloor n/2 \rfloor }  \big (\cD_{\a(2)} \big )^j \f^{\perp}_{\a(n-2j)}~. 
\eea
Here each of the fields $\f^{\perp}_{\a(n-2j)}$ are transverse, except of course $\phi^{\perp}$ and $\f^{\perp}_{\a}$. We note that, using \eqref{splitting}, one may take the decomposition \eqref{Decomp} a step further and bisect each term into irreducible components which are transverse and have positive or negative helicity,
\begin{align}
\f_{\a(n)} = \sum_{j=0}^{\lfloor n/2 \rfloor }  \big (\cD_{\a(2)} \big )^j\Big( \f^{(+)}_{\a(n-2j)}+\f^{(-)}_{\a(n-2j)}\Big)~. 
\end{align}

Making use of the projectors \eqref{SimpBosProj} and \eqref{SimpFermProj} and their corresponding properties, one can construct operators which extract the component $\phi_{\a(n-2j)}^{\perp}$ from the decomposition \eqref{Decomp}, where $1\leq j \leq \lfloor n/2 \rfloor$. In particular, we find that the spin $\frac{1}{2}(n-2j)$ component may be extracted via
\begin{align}
\phi_{\a(n)}\mapsto \phi^{\perp}_{\a(n-2j)}=\big(\mb{S}_{[n-2j]}^{\perp}\phi\big)_{\a(n-2j)}\equiv \mb{S}_{\a(n-2j)}^{\perp}(\phi)~,
\end{align}
where we have defined
\begin{align} \label{Extractors}
\mb{S}_{\a(n-2j)}^{\perp}(\phi)&=\frac{(-1)^j}{2^{2j}}\binom{n}{j} \prod_{k=1}^{j}\big (\cQ - \t_{(k,n-2j+2k)}\cS^2 \big )^{-1} \P^{\perp}_{[n-2j]}\big(\cD^{\b(2)}\big)^j\f_{\a(n-2j)\b(2j)}~.
\end{align}
From this expression, it is clear that $\mb{S}_{\a(n-2j)}^{\perp}(\phi)$ is transverse,
\be
0=\cD^{\b(2)}\mb{S}_{\b(2)\a(n-2j-2)}^{\perp}(\phi)~.
\ee
Therefore it is appropriate to call $\mb{S}_{[n-2j]}^{\perp}$ the transverse spin $\frac{1}{2}(n-2j)$ extractor. It is not a projector, since it is dimensionful and reduces the rank of the field on which it acts. 

Let $\psi_{\a(n)}$ be some longitudinal field, 
$\psi_{\a(n)}=\mc{D}_{\a(2)}\zeta_{\a(n-2)}$. 
We do not assume it to be in the image of $\Pi^{\parallel}_{[n]}$. 
However, since $\P_{[n]}^{\perp}$ commutes with $\mc{D}_{\a(2)}$ and annihilates all lower-rank fields, eq. \eqref{BosProjProp}, it follows that it also annihilates any rank-$n$ longitudinal field\footnote{This also implies that $\widehat{\P}^{\perp}_{[n]}\psi_{\a(n)}=0$, since both $\widehat{\P}^{\perp}_{[n]}$ and $\P^{\perp}_{[n]}$ are equal on $\mc{V}_{(n)}$. }
\begin{align}\label{LongitudinalKiller}
\psi_{\a(n)}=\mc{D}_{\a(2)}\z_{\a(n-2)}\qquad \implies \qquad \P^{\perp}_{[n]}\psi_{\a(n)}=0~.
\end{align}
As a consequence, given two integers $m,n$ satisfying $2\leq m \leq n$, it immediately follows that $\Pi^{\parallel}_{[n]}$ acts as the identity operator on the space of rank-$m$ longitudinal fields $\psi_{\a(m)}$,
\begin{align} \label{Transkillslong}
\psi_{\a(m)}=\mc{D}_{\a(2)}\psi_{\a(m-2)} \qquad \implies \qquad \Pi^{\parallel}_{[m+2s]}\psi_{\a(m)}=\psi_{\a(m)}~,
\end{align}
with $s$ a non-negative integer.
These properties will be useful in section \ref{secCT}.

Decompositions similar to \eqref{2.51} are well-known in the literature (usually they are stated without a derivation) and are used in the framework of path-integral quantisation, see e.g. \cite{GGS}. Making use of the projectors allows one to reconstruct
$\f^{\perp}_{\a(n)} $ and $\f_{\a(n-2)} $ from $\f_{\a(n)} $.
Quite often such decompositions are given in vector notation  in terms of a symmetric  field $\vf_{a_1 \dots a_s} =\vf_{(a_1 \dots a_s )}$ subject to  the double traceless constraint $\vf_{a_1 \dots a_{s-4} bc}{}^{bc}=0$ (Fronsdal's field \cite{Fronsdal2}). The decomposition in AdS$_3$ reads \cite{GGS} 
\bea \label{DoubleTdecomp}
\vf_{a_1 \dots a_s} = \vf^{\rm TT}_{a_1 \dots a_s} + \eta_{(a_1a_2 } \widetilde{\vf}_{a_3 \dots a_s)}
+\cD_{ (a_1} \z_{a_2 \dots a_s)} ~, \qquad \cD^b \vf^{\rm TT}_{b a_1 \dots a_{s-1} } =0~,
\label{2.59}
\eea
where $\vf^{\rm TT}_{a_1 \dots a_s}$, $\widetilde{\vf}_{a_1 \dots a_{s-2} }$ and 
$\z_{a_1 \dots a_{s-1} }$ are symmetric and traceless. This decomposition for a symmetric 
second-rank tensor field, $\vf_{ab}=\vf_{ba}$,  in a curved four-dimensional space was introduced long ago 
\cite{Deser67,York73,York74,GibbonsPerry}. In this paper we consider only symmetric traceless fields $\vf_{a_1 \dots a_s}$ satisfying the constraint 
 $\vf_{a_1 \dots a_{s-2}b}{}^b= 0$. 
 In this case, $\widetilde{\vf}_{a_1 \dots a_{s-2}}$ in the 
 decomposition \eqref{2.59}  is given by 
\bea
\widetilde{\vf}_{a_1 \dots a_{s-2}} =  - \frac{s-1}{2s-1}\cD^b \z_{a_1 \dots a_{s-2}b}~.
\eea


\subsection{Linearised higher-spin Cotton tensors} \label{secCT}

Further applications of spin projection operators can be found in modern conformal higher-spin theories. In particular, we will show that the spin projectors can be used to obtain new realisations of the linearised higher-spin Cotton tensors, which were recently derived in \cite{KP1}. For integer $n\geq 2$, the higher-spin bosonic and fermionic Cotton tensors $\mathfrak{C}_{\a(n)}(h)$  take the respective closed forms
\begin{subequations}\label{cotE}
	\begin{align}
	\mathfrak{C}_{\a(2s)}(h)&=\frac{1}{2^{2s-1}}\sum_{j=0}^{s-1}2^{2j+1}\binom{s+j}{2j+1}\prod_{t=1}^{j}\Big(\cQ-\tau_{(s-t,2s)}\cS^2\Big) \non\\
	&\phantom{\frac{1}{2^{2s-1}}\sum_{j=0}^{s-1}2^{2j+1}\binom{s+j}{2j+1}}\times
	\cD_{\a(2)}^{s-j-1}\cD_{\a}{}^{\b}\big(\cD^{\b(2)}\big)^{s-j-1}h_{\a(2j+1)\b(2s-2j-1)}~, \label{cotF}\\
	\mathfrak{C}_{\a(2s+1)}(h)&=\frac{1}{2^{2s}}\sum_{j=0}^{s}2^{2j}\binom{s+j}{2j}\frac{(2s+1)}{(2j+1)}\prod_{t=1}^{j}\Big(\cQ-\tau_{(s-t+1,2s+1)}\cS^2\Big) ~~~~~~~~~~~~~~~~~~~~~~~~~~~\non\\
	&\phantom{\frac{1}{2^{2s}}\sum_{j=0}^{s}\binom{s+j}{2j}\frac{(2s+1)}{(2j+1)}}\times \cD_{\a(2)}^{s-j}\big(\cD^{\b(2)}\big)^{s-j}h_{\a(2j+1)\b(2s-2j)}~.\label{cotB}
	\end{align}
\end{subequations}
The Cotton tensors are primary descendents of  the conformal gauge field $h_{\a(n)}$, which is a real field defined modulo gauge transformations of the form
\be
\d_\z h_{\a(n)} = \cD_{\a(2)}\z_{\a(n-2)}~,
\ee
for some real unconstrained gauge parameter $\z_{\a(n-2)}$. The Cotton tensors \eqref{cotE} are characterised by the properties:
\begin{enumerate}
	\item $\mathfrak{C}_{\a(n)}(h)$ is transverse
	\bsubeq \label{CotProp}
	\be \label{CTP1}
	\cD^{\b\g}\mathfrak{C}_{\b\g\a(n-2)}(h)=0~.
	\ee
	\item $\mathfrak{C}_{\a(n)}(h)$  is gauge-invariant
	\be \label{CTP2}
	\mathfrak{C}_{\a(n)}(\d_\z h) = 0~.
	\ee
	\esubeq
\end{enumerate}
Making use of the bosonic \eqref{BP} and fermionic \eqref{FP} spin projectors $\widehat{\Pi}^{\perp}_{[n]}$, we see that the higher-spin Cotton tensors \eqref{cotE} can be recast into the simple form:
\begin{subequations} \label{CT}
	\bea
	\mathfrak{C}_{\a(2s)} (h) &=& \frac{1}{2s} \prod_{t=1}^{s-1}\big (\mathcal{Q} - \t_{(t,2s)} \mathcal{S}^2 \big ) \mathcal{F} \widehat{\Pi}^{\perp}_{[2s]}h_{\a(2s)}~, \label{CTB} \\
	\mathfrak{C}_{\a(2s+1)}(h) &=&\prod_{t=1}^{s}\big (\mathcal{Q} - \t_{(t,2s+1)} \mathcal{S}^2 \big )\widehat{\Pi}^{\perp}_{[2s+1]}h_{\a(2s+1)}~.\label{CTF}
	\eea 
\end{subequations}
The identity $\cF \cD_{(-2)}^s \f_{\a(2s)} = 0$ proves useful in deriving \eqref{CTB}. In the flat space limit, $\cS \rightarrow 0$, \eqref{CT} coincides with the closed form expressions of $\mf{C}_{\a(n)}(h)$ given in \cite{K,PT}.\footnote{It can be shown that the Cotton tensors are equivalent to those derived in \cite{HHL, HHL2}.} Moreover, we can make use of the equivalent family of projectors $\P^{\perp}_{[n]}$ to recast $\mathfrak{C}_{\a(n)}(h)$ purely in terms of the quadratic Casimir operators \eqref{QC}. Explicitly, they read
\begin{subequations} \label{CT1}
	\bea
	\mathfrak{C}_{\a(2s)} (h) &=& \frac{\cF}{2^{2s-1}(2s-1)!}  \prod_{j=1}^{s-1} \Big ( \cF^2 -4j^2 \big (\cQ-4(j-1)(j+1)\cS^2 \big ) \Big )   h_{\a(2s)}~,  \\
	\mathfrak{C}_{\a(2s+1)}(h) &=&\frac{1}{2^{2s}(2s)!}\prod_{j=0}^{s-1} \Big ( \cF^2 -(2j+1)^2 \big (\cQ-(2j-1)(2j+3)\cS^2 \big ) \Big ) h_{\a(2s+1)}~.
	\eea
\end{subequations}

There are many advantages to expressing the Cotton tensors in terms of spin projection operators. Firstly, in both \eqref{CT} and \eqref{CT1}, the properties of (i) transversality \eqref{CTP1} and (ii) gauge invariance \eqref{CTP2} are manifest, as a consequence of the projector properties \eqref{ProjTrans}  and \eqref{LongitudinalKiller} respectively. Using this gauge freedom,  one may impose the transverse gauge condition on $h_{\a(n)}$, 
\begin{align}
h_{\a(n)}\equiv h^{\rm T}_{\a(n)}~, \qquad 0=\mc{D}^{\b(2)}h^{\rm T}_{\b(2)\a(n-2)}~.
\end{align}
  On account of \eqref{ProjU}, in this gauge the Cotton tensors become manifestly factorised into products of second order differential operators involving all partial masses,
  \bsubeq
  \bea
  \mf{C}_{\a(2s)}(h^{\rm T}) &=&  \frac{1}{2s} \prod_{t=1}^{s-1}\big (\mathcal{Q} - \t_{(t,2s)} \mathcal{S}^2 \big ) \mathcal{F} h^{\rm T}_{\a(2s)} ~, \\
  \mathfrak{C}_{\a(2s+1)}(h^{\rm T}) &=&\prod_{t=1}^{s}\big (\mathcal{Q} - \t_{(t,2s+1)} \mathcal{S}^2 \big )h^{\rm T}_{\a(2s+1)}~.
  \eea
\esubeq
  This property was observed in \cite{KP1} without the use of projectors.
  An interesting feature of the new realisation \eqref{CT1}, which was not observed in \cite{KP1}, is that the Cotton tensors are manifestly factorised in terms of second-order differential operators without having to enter the transverse gauge.  

By virtue of the above observations, it follows that the conformal higher-spin action \cite{KP2, KP3}
\begin{align} 
S_{\text{CHS}}^{(n)}[h]=\frac{\text{i}^n}{2^{\lceil n/2 \rceil+1}}\int\text{d}^3x\, e \, h^{\a(n)}\mf{C}_{\a(n)}(h)
\label{CSA}
\end{align}
is manifestly gauge invariant and factorised when $\mf{C}_{\a(n)}(h)$ is expressed as in \eqref{CT1}.

Analogous factorised expressions can be given for 
the so-called new topologically massive (NTM) models.
For  bosonic fields they were first introduced in \cite{BKRTY} in Minkowski space. Extensions of these models to fields with half-integer spin were proposed in \cite{KP2}, where their generalisations to an AdS background were also given. 
These models are formulated solely in terms of the gauge prepotentials $h_{\a(n)}$ and 
the associated Cotton tensors $\mf{C}_{\a(n)}(h)$.  
Given an integer $n\geq 2$, the gauge-invariant NTM action for the field $h_{\a(n)}$ 
given in \cite{KP2} is  
\begin{align}
S_{\text{NTM}}^{(n)}[h]=\frac{\text{i}^n}{2^{\lceil n/2 \rceil+1}}\frac{1}{\rho}\int\text{d}^3x\, e \, h^{\a(n)}
 \big(\mc{F}-\s\rho\big) \mf{C}_{\a(n)}(h)
~, \label{HSNTMG}
\end{align}
where $\rho$ is some positive mass parameter and $\s:=\pm 1$. Making use of  the representation \eqref{CT1} leads to a manifestly gauge invariant and factorised form for the action \eqref{HSNTMG}. The equation of motion obtained by varying \eqref{HSNTMG} with respect to the field $h^{\a(n)}$ is 
\begin{align}
0=\big(\mc{F}-\s \rho\big)\mf{C}_{\a(n)}(h)~. \label{EOM1}
\end{align}
By analysing \eqref{EOM1}, it can be shown that on-shell, the  action \eqref{HSNTMG} describes a propagating mode with pseudo-mass $\r$, spin $n/2$ and helicity $\s n/2$ given $\r \neq \r_{(t,2s)}$. For the case $\r = \r_{(t,2s)}$, the model describes only pure gauge degrees of freedom.

Recently, a new variant of the NTM model for bosonic fields in $\mb{M}^3$ was proposed in \cite{DS}. This model also does not require auxilliary fields, but is of order $2s-1$ in derivatives, whereas those given in \cite{BKRTY} are of order $2s$. Given an integer $s\geq 1$,  the actions of \cite{DS} may be readily extended to AdS$_3$ as follows
\bea \label{NNTM}
\widetilde{S}_{\text{NTM}}^{(2s)}[h] =\int \rd^3 x\, e\,  h^{\a(2s)} \big ( \cF - \s \r \big ) \mf{W}_{\a(2s)}(h)  ~,
\eea
where $\r$ is a positive mass parameter, $\s:=\pm 1$, and  $\mf{W}_{\a(2s)}(h)$ is the field strength,
\be
\mf{W}_{\a(2s)}(h):=\prod_{t=1}^{s-1}\big (\mathcal{Q} - \t_{(t,2s)} \mathcal{S}^2 \big )  {\Pi}^{\perp}_{[2s]}h_{\a(2s)}~.
\ee
Due to the properties of $\Pi^{\perp}_{[2s]}$, the action \eqref{NNTM} is manifestly gauge invariant and factorised.  The descendent $\mf{W}_{\a(2s)}(h)$ may be obtained from $\mf{C}_{\a(2s)}(h)$ by stripping off $\mc{F}$: 
\begin{align}
\mf{C}_{\a(2s)}(h)=\frac{1}{2s}\mc{F}\mf{W}_{\a(2s)}(h)~.
\end{align}
 A similar construction does not appear to be possible in the fermionic case. 

The equation of motion obtained by varying \eqref{NNTM} with respect to the field $h^{\a(2s)}$ is
\bea \label{EoM}
0 = (\cF - \s \r)\mf{W}_{\a(2s)}(h)~.
\eea
By analysing \eqref{EoM}, it can be shown that on-shell, the model \eqref{NNTM} has the same particle content as the NTM model \eqref{HSNTMG}.


\subsection{Results in Minkowski space} \label{secProjM}

In this section we study the flat-space limit of various results derived in section \ref{section2}. Of particular interest are the transverse projectors  which are  constructed in terms of the Casimir operators of  $\mathfrak{so}(2,2)$. In this limit we obtain novel realisations for the transverse projectors on $\mb{M}^{3}$ which did not  appear in \cite{IP,BLFKP}. They are expressed in terms of the quadratic Casimir operators of the three dimensional Poincar\'e algebra $\mathfrak{iso}(2,1)$,  
\bsubeq
\begin{alignat}{2}  \label{QCMS}
\Box&:=\pa^a\pa_a = -\hf \pa^{\a\b}\pa_{\a\b}~, \\
\mc{W}&:= \pa^{\a\b}M_{\a\b}~,  &[\cW, \partial_{\a\b}] = 0~.
\label{2.62b}
\end{alignat}
\esubeq
Here $\pa_{\a\b}$ are the partial derivatives of $\mb{M}^3$ and
 $\mc{W}$ is the Pauli-Lubanski pseudo-scalar. We recall that an irreducible representation of $\mathfrak{iso}(2,1)$ with mass $\rho$ and helicity $\s n/2$ may be realised on the space of totally symmetric rank-$n$ spinor fields $\phi_{\a(n)}$ satisfying the differential equations
 \begin{subequations}
 \begin{align}
 \pa^{\b(2)}\f_{\b(2)\a(n-2)}&=0~,\\
 \big( \mc{W}-\s n \rho \big)\phi_{\a(n)}&=0~, \label{irredMink3b}
 \end{align} 
 \end{subequations}
where $\s=\pm 1$. These equations are equivalent to those given in \cite{GKL, TV}. We are concerned only with representations carrying (half-)integer spin.
 
 By taking the limit $\cS \rightarrow 0$ of the corresponding AdS$_3$ expressions given above, one may obtain the following results in Minkowski space:
 \begin{itemize}
\item  The bosonic \eqref{SimpBosProj} and fermionic \eqref{SimpFermProj} spin projection operators reduce to
\bsubeq \label{3dSPinproj}
\bea 
\cP^{\perp}_{[2s]}
&=& \frac{1}{2^{2s-1}(2s)!  \Box^s }  \prod_{j=0}^{s-1} \Big ( \cW^2 -(2j)^2 \Box \Big )  
~, \\
\cP^{\perp}_{[2s+1]} 
&=& \frac{1}{2^{2s}(2s)!\Box^s}\prod_{j=0}^{s-1} \Big ( \cW^2 -(2j+1)^2 \Box \Big ) 
~.
\eea
\esubeq

\item The orthogonal helicity projectors \eqref{helicityproj} reduce to
\be \label{FSHP}
\mathds{P}^{(\pm)}_{[n]} =\hf \bigg (\mathds{1} \pm \frac{\cW}{n\sqrt{\Box}} \bigg ) \cP^{\perp}_{[n]}~.
\ee 
From \eqref{helcityprojprop} it follows that the field $\f_{\a(n)}^{(\pm)}:=\mathds{P}^{(\pm)}_{[n]} \f_{\a(n)} $ satisfies
\be
\big ( \cW \mp  n \sqrt{\Box} \big )\f_{\a(n)}^{(\pm)} = 0 ~.
\ee
For a $\phi_{\a(n)}$ lying on the mass shell, $\big(\Box-\rho^2\big)\phi_{\a(n)}=0$, this reduces to \eqref{irredMink3b}.
\item The transverse spin $\hf(n-2j)$ extractors \eqref{Extractors}, where $1\leq j\leq \lfloor n/2 \rfloor$, are given by
\begin{align}
\mathds{S}_{\a(n-2j)}^{\perp}(\phi)&=\frac{(-1)^j}{2^{2j}}\binom{n}{j} \frac{1}{\Box^j} \cP^{\perp}_{[n-2j]}\big(\partial^{\b(2)}\big)^j\f_{\a(n-2j)\b(2j)}~.
\end{align}

\item The new realisations for the higher-spin Cotton tensors \eqref{CT1}  become
\begin{subequations} \label{CT2}
	\bea
	\cC_{\a(2s)} (h) &=& \frac{\cW}{2^{2s-1}(2s-1)!}  \prod_{j=1}^{s-1} \Big (\cW^2 -(2j)^2\Box \Big )   h_{\a(2s)}~, \\
	\cC_{\a(2s+1)}(h) &=&\frac{1}{2^{2s}(2s)!}\prod_{j=0}^{s-1} \Big ( \cW^2 -(2j+1)^2 \Box  \Big ) h_{\a(2s+1)}~. 
	\eea
\end{subequations}
\end{itemize}

It may be shown that each of these expressions are equivalent to the corresponding ones given in \cite{BLFKP}, except for the lower-spin extractors, which were not discussed in \cite{BLFKP}.


\section{Transverse superprojectors in AdS$^{3|2}$} \label{secSP}

In this section, we derive the superprojectors in $\cN=1$ AdS superspace, AdS$^{3|2}$, and explore several of their applications. We remind the reader that AdS$^{3|2}$ is the maximally supersymmetric solution of three-dimensional $\cN=1$ AdS supergravity \cite{GGRS}.

We begin by reviewing the geometric structure of AdS$^{3|2}$, as presented in \cite{KLTM}, 
which is described in terms of its covariant derivatives\footnote{In the hope that no confusion arises, we use the same notation for the vector covariant derivative in AdS$_3$ and in AdS$^{3|2}$.
}
\be \label{SCD}
\cD_A = (\cD_a, \cD_\a) = E_A{}^M {\partial}_M + \hf \O_A{}^{bc}M_{bc}~. 
\ee
Here $E_A{}^M$ is the inverse supervielbein and  $\O_A{}^{bc}$ the Lorentz connection. The covariant derivatives obey the following (anti-)commutation relations\footnote{In vector notation, the commutation relations \eqref{SA2} take the form $ [ \cD_a, \cD_\b ] = \cS (\g_a)_\b{}^\g \cD_\g$ and  $ [ \cD_a, \cD_b ] = -4 \cS^2 M_{ab}$.}
\begin{subequations} \label{SA}
	\be
	\{ \cD_\a , \cD_\b \} = 2\ri \cD_{\a\b} - 4\ri\cS M_{\a\b}~, \label{SA1} \\
	\ee
	\be
	\ [ \cD_{\a \b}, \cD_\g ] = -2\cS \ve_{\g(\a}\cD_{\b)}~, \qquad \ [ \cD_{\a \b}, \cD_{\g \d} ] = 4 \cS^2 \Big(\ve_{\g(\a}M_{\b)\d} + \ve_{\d(\a} M_{\b)\g}\Big)~, \label{SA2}
	\ee
\end{subequations}
where $\cS\neq 0$ is a real constant parameter which determines the curvature of AdS$^{3|2}$. 

We list several identities which prove indispensable for calculations:
\begin{subequations}  \label{CDI}
	\bea 
	\cD_\a \cD_\b &=& \ri \cD_{\a\b} - 2\ri\cS M_{\a\b}+\frac{1}{2}\ve_{\a\b}\cD^2~, \\
	\cD^{\b} \cD_\a \cD_\b &=& 4\ri \cS\cD_\a~, \quad \{ \cD^2, \cD_\a  \} = 4\ri \cS\cD_\a~,\\
	\cD^2 \cD_\a &=& 2\ri \cS\cD_\a + 2\ri \cD_{\a\b} \cD^\b - 4\ri \cS \cD^\b M_{\a\b}~, \\
	\qquad \ [ \cD_{\a} \cD_{\b}, \cD^2 ]
	&=& 0 \quad \Longrightarrow \quad \ [\cD_{\a\b}, \cD^2 ] = 0~,
	\eea
\end{subequations}
where we have denoted $\cD^2 = \cD^\a \cD_\a$. These relations can be derived from the algebra of covariant derivatives \eqref{SA}.

Crucial to our analysis are two independent Casimir operators of the $\cN=1$ AdS$_3$ isometry supergroup $\text{OSp}(1|2;\mathbb{R}) \times \text{SL}(2,\mathbb{R})$. They are  \cite{KP1,KP2} 
\bsubeq \label{SCQ}
\begin{alignat}{2}
\mathbb{Q}:&=- \frac{1}{4}\cD^2 \cD^2 + \ri \cS \cD^2~, \qquad &[\mathbb{Q}, \cD_A]= 0~, \label{QQC}\\
\mathbb{F}:&=- \frac{\ri}{2}\cD^2 + 2\cD^{\a\b}M_{\a\b}~,  &[\mathbb{F}, \cD_A]= 0~. \label{FQC}
\end{alignat}
\esubeq
Making use of the identity
\bea
-\frac{1}{4}\cD^2\cD^2 &=&  \Box - 2\ri\cS \cD^2+2\cS \cD^{\a\b}M_{\a\b} -2\cS^2M^{\a\b}M_{\a\b}~,
\eea
allows us to express $\mathbb{Q}$ in terms of the d'Alembert operator $\Box = \cD^a \cD_a $.
The operators $\mathbb{Q}$ and $\mathbb{F}$ are related to each other as follows
\bea \label{FSquared}
\mb{F}^2\F_{\a(n)} &=& \Big ( (2n +1)^2\mb{Q}  + (2n+1)(2n^2+2n-1)\ri \cS \cD^2 +4n^2(n+2)^2 \cS^2\Big  ) \F_{\a(n)}  \non \\
&&+ 4 (2n^2+n-2)\ri \cS \cD_{\a}\cD^\b \F_{\b \a(n-1)} -4\ri n\cD_{\a\b}\cD^{\b} \cD^{\g}\F_{\g\a(n-1)}~ \non \\
&&+4n(n-1)\cD_{\a(2)}\cD^{\b(2)}\F_{\b(2)\a(n-2)}~,
\eea
 for an arbitrary symmetric rank-$n$ spinor superfield  $\Phi_{\a(n)}$.


\subsection{On-shell superfields}
We begin by reviewing aspects of on-shell superfields in AdS$^{3|2}$, as presented in \cite{KP1}.
Given an integer $n \geq 1$, the real symmetric superfield $\F_{\a(n)}$ is said to be on-shell if it satisfies the two constraints 
\bsubeq \label{SOC}
\bea
0 &=& \cD^\b \F_{\b\a(n-1)}~, \label{STrans} \\
0 &=& \big ( \mathbb{F} - \s M \big ) \F_{\a(n)}~, \label{SFO}
\eea
\esubeq
where $\s:=\pm 1$ and $M\geq 0$ is a real parameter of unit mass dimension. 
Such a field furnishes an irreducible representation of the $\cN=1$ AdS$_3$ superalgebra $\mathfrak{osp}(1|2;{\mathbb R} ) \oplus \mathfrak{sl}(2, {\mathbb R})$, which we denote as 
$\mathfrak{S}(M, \s \frac{n}{2})$. It can be shown that the representation $\mathfrak{S}(M, \s \frac{n}{2})$ decomposes into two irreducible representations of $\mathfrak{so}(2,2)$,
\be \label{decomp}
\mathfrak{S} \Big (M, \s \frac{n}{2} \Big ) = \mathfrak{D} \Big ( \r_A, \s_A \frac{n}{2} \Big ) \oplus \mathfrak{D} \Big ( \r_B, \s_B\frac{n+1}{2} \Big )~.
\ee
Here, the pseudo-masses are given by
\bea
\r_A= \frac{n}{2n+1}\Big | \s M-(n+2)\cS \Big |~, \qquad
\r_B= \frac{n+1}{2n+1} \Big | \s M + (n-1)\cS \Big |~,
\eea
and the corresponding signs of the superhelicities are  
\bea
\s_A &=& \frac{  \s M-(n+2)\cS }{\big | \s M-(n+2)\cS \big |}~, \qquad
\s_B = \frac{ \s M + (n-1)\cS}{\big | \s M + (n-1)\cS \big |}~.
\eea

The representation $\mathfrak{S}(M, \s \frac{n}{2})$ is unitary if the parameter $M$ obeys the unitarity bound $M\geq 2(n-1)(n+1)\cS$. This bound ensures that both representations appearing in the decomposition \eqref{decomp} are unitary.

A superfield satisfying the first condition \eqref{STrans} is said to be transverse. Any transverse superfield may be shown to satisfy the following relation
\be \label{Transprop1}
-\frac{\ri}{2}\cD^2\F_{\a(n)} = \cD_{(\a_1}{}^\b\F_{\a_2 ... \a_n)\b} + (n+2)\cS \F_{\a(n)}~.
\ee
If a transverse superfield also satisfies \eqref{SFO}, we say that it carries pseudo-mass $M$, superspin $n/2$ and superhelicity $\hf(n+\hf)\s$. From \eqref{Transprop1} it follows that an on-shell superfield \eqref{SOC} satisfies
\be \label{FOSM}
-\frac{\ri}{2}\cD^2\F_{\a(n)} = \frac{1}{2n+1}\Big (\s M+2n(n+2)\cS \Big ) \F_{\a(n)}~,
\ee
and hence  the second-order mass-shell equation
\bsubeq \label{SOMSOM}
\begin{align} \label{SOMS}
0 &= \big ( \dsQ -  \l^2\big ) \F_{\a(n)}~, \\
 \l^2:=\frac{1}{(2n+1)^2} \big [\s M +2n&(n+2)\cS \big ] \big [\s M+2(n-1)(n+1)\cS \big ]~.
\end{align}
\esubeq
The equations \eqref{STrans} and \eqref{FOSM} were introduced in \cite{KNT-M}. On the other hand, one may instead consider a superfield $\F_{\a(n)} $ satisfying \eqref{STrans} and \eqref{SOMS}. In this case, using the identity \eqref{FSquared}, one can show that \eqref{SOMS} becomes
\bea
0=\Big ( \mb{F} - \s_{(-)}| M_{(-)}|\Big ) \Big ( \mb{F} - \s_{(+)}|M_{(+)}| \Big )~,
\eea
where we have defined $\s_{(\pm)}=\text{sgn}(M_{(\pm)}) $ and
\bea
M_{(\pm)} := -(2n^2+2n-1)\cS \pm (2n+1)\sqrt{\l^2+\cS^2} ~.
\eea
It follows that such a field furnishes the reducible representation 
\be \label{redS}
 \mf{S} \Big (|M_{(-)}|, \s_{(-)} \frac{n}{2} \Big ) \oplus \mf{S} \Big (|M_{(+)}|, \s_{(+)}\frac{n}{2} \Big )~.
\ee

In AdS$^{3|2}$ there exist two distinct types of on-shell partially massless superfields \cite{KP1}, which are distinguished by the sign $\s$ of their superhelicity.   More specifically, they are described by an on-shell superfield \eqref{SOC} whose pseudo-mass and parameter $\s$ assume the special combinations
\bsubeq \label{PsudeoM}
\begin{align}
\s=+1~,\qquad M &\equiv M^{(+)}_{(t,n)} = 2 \big [ n(n-2t+1)-(t-1) \big ] \cS~,  &1&\leq t \leq \lfloor n/2 \rfloor ~, \\
\s=-1~,\qquad M &\equiv M^{(-)}_{(t,n)} = 2\big [n(n-2t)-(t+1) \big ] \cS ~,  &0&\leq t \leq \lceil n/2 \rceil -1~. \label{typeAPM}
\end{align}
\esubeq
The integer $t$ is called the (super)depth and the corresponding supermultiplets are denoted by $\Phi^{(t,+)}_{\a(n)}$ and $\Phi^{(t,-)}_{\a(n)}$ respectively. Their second order equations \eqref{SOMSOM} take the form
\be
0 = \big ( \dsQ - \l_{(t,n)}^{(+)} \cS^2 \big ) \F_{\a(n)}^{(t,+)}~, \quad 0= \big ( \dsQ - \l_{(t,n)}^{(-)} \cS^2 \big ) \F_{\a(n)}^{(t,-)}~,
\ee
where we have introduced the partially massless values
\be
\l_{(t,n)}^{(+)}=4(n-t)(n-t+1)~, \qquad \l_{(t,n)}^{(-)} = 4t(t+1)~.
\ee

The gauge symmetry associated with positive and negative superhelicity partially massless superfields of depth-$t$ is
\begin{subequations} \label{SPMG}
\begin{align}
\delta_{\L}\F^{(t,+)}_{\a(n)}&=
\phantom{\rm{i}^n} \big(\mc{D}_{\a(2)}\big)^t \L_{\a(n-2t)}~, &&1 \leq t \leq \lfloor n/2 \rfloor~, \label{SPMG2} \\
\delta_{\L}\F^{(t,-)}_{\a(n)}&=
\text{i}^n \big(\mc{D}_{\a(2)}\big)^t\mc{D}_{\a}\L_{\a(n-2t-1)}~, &&0 \leq t \leq \lceil n/2 \rceil - 1~.  \label{SPMG1}
\end{align}
\end{subequations}
In particular, the system of equations \eqref{SOC} and \eqref{PsudeoM} is invariant under these transformations for an on-shell real gauge parameter. 


\subsection{Superspin projection operators}

We wish to find supersymmetric generalisations of 
the spin projection operators in AdS$_3$ which were computed in section \ref{section2}. 
More precisely, let us denote by $\mathds{V}_{(n)}$ the space of totally symmetric rank-$n$ superfields $\F_{\a(n)} $ 
on AdS$^{3|2}$. For any integer $n\geq 1$, we define the 
rank-$n$ superspin projection operator\footnote{The four-dimensional analogue was recently given in \cite{BHKP}.}  $\bm\P^{\perp}_{[n]}$ to act on $\mathds{V}_{(n)}$ 
by the rule
\bea 
\bm \Pi^{\perp}_{[n]}: \mathds{V}_{(n)} \longrightarrow \mathds{V}_{(n)}~, \qquad \F_{\a(n)} \longmapsto  \bm \Pi^{\perp}_{[n]} \F_{\a(n)}~  =:\F^{\perp}_{\a(n)}~,
\eea
which satisfies the following properties:
\begin{enumerate} 
	\item  $\bm \Pi^{\perp}_{[n]}$ is idempotent, 
	\bsubeq  \label{SProp}
	\be 
	\bm \Pi^{\perp}_{[n]}\bm \Pi^{\perp}_{[n]}=\bm \Pi^{\perp}_{[n]}~.
	\ee
	\item  $\bm \Pi^{\perp}_{[n]}$ maps $\F_{\a(n)}$ to a transverse superfield,
	\be 
	\cD^{\b}\F^{\perp}_{\b\a(n-1)} =0~.
	\ee
	\item Every transverse superfield  $\J_{\a(n)}$  belongs to the image of $\bm \Pi^{\perp}_{[n]}$,
	\be \label{SIT}
	\mc{D}^{\b}\J_{\b\a(n-1)}=0~\quad\implies\quad\bm \Pi^{\perp}_{[n]} \J_{\a(n)} = \J_{\a(n)}~.
	\ee
	\esubeq
\end{enumerate}
In other words, the superprojector $\bm \Pi^{\perp}_{[n]}$ maps $\F_{\a(n)}$ to a supermultiplet with the properties of a conserved supercurrent.

To obtain a superprojector, we introduce the operator $\D^{\a}{}_\b$ \cite{KP2}
\be \label{TransOp}
\D^{\a}{}_\b:=-\frac{\ri}{2}\cD^\a \cD_\b - 2\cS \d^{\a}{}_{\b}~, \quad \cD^\b \D^{\a}{}_\b =  \D^{\a}{}_\b \cD_\a = 0~,
\ee
and its corresponding extensions \cite{KP1} 
\be \label{ExTransOp}
\D^{\a}_{[j]}{}_\b := -\frac{\ri}{2}\cD^\a \cD_\b - 2j\cS \d^{\a}{}_{\b}~. 
\ee
Note that for the case $j=1$, \eqref{ExTransOp} coincides with \eqref{TransOp}. It can be shown that the operator \eqref{ExTransOp} has the following properties
\bsubeq \label{Ident}
\bea
[\D^{\a_1}_{[j]}{}_{\b_1},\D^{\a_2}_{[k]}{}_{\b_2}] &=& \varepsilon_{\b_1 \b_2}\cS \big ( \cD^{\a(2)}-\cS M^{\a(2)} \big ) - \varepsilon^{\a_1 \a_2}\cS \big ( \cD_{\b(2)}-\cS M_{\b(2)} \big ) ~,  \label{Ident1}\\
\varepsilon^{\b_1 \b_2}\D^{\a_1}_{[j]}{}_{\b_1}\D^{\a_2}_{[j+1]}{}_{\b_2}&=&-j  \varepsilon^{\a_1 \a_2}\cS \big (\ri \cD^2 + 4(j+1)\cS^2 \big )~, \label{Ident2} \\
\varepsilon_{\a_1 \a_2}\D^{\a_1}_{[j+1]}{}_{\b_1}\D^{\a_2}_{[j]}{}_{\b_2}&=& j  \varepsilon_{\b_1 \b_2}\cS \big (\ri \cD^2 + 4(j+1)\cS^2 \big )~, \label{Ident3}\\
\D^{\b}_{[j]}{}_{\a}\D^{\g}_{[k]}{}_{\b}&=& - \frac{\ri}{2}\cD^2 \D^{\g}_{[1]}{}_{\a} +  (j+k-1) \ri  \cS  \cD^\g \cD_\a + 4jk\cS^2 \d_\a{}^\g ~,\label{Ident4}  \\
\ [ \D^{\a}_{[j]}{}_\b, \cD^2 ] &=& 0~,
\eea
\esubeq
for arbitrary integers $j$ and $k$. 

Let us define the operator $\mathbb{T}_{[n]}$, which acts on $\mathds{V}_{(n)}$ by the rule
\be \label{TSO}
\mathbb{T}_{[n]} \F_{\a(n)} \equiv \mathbb{T}_{\a(n)}(\F) =\D^{\b_1}_{[1]}{}_{(\a_1}\D^{\b_2}_{[2]}{}_{\a_2} \cdots \D^{\b_n}_{[n]}{}_{\a_n)}\F_{\b(n)}~. 
\ee
This operator maps $\F_{\a(n)}$ to a transverse superfield
\be \label{STransprop}
\cD^\b \mathbb{T}_{\b\a(n-1)}( \F)=0~.
\ee
To see this, one needs to open the symmetrisation in \eqref{TSO}
\bea
\cD^\b \mathbb{T}_{\b\a(n-1)}( \F) &=&  \cD^\g\D^{\b_1}_{[1]}{}_{(\g}\D^{\b_2}_{[2]}{}_{\a_1} \cdots \D^{\b_n}_{[n]}{}_{\a_{n-1})}\F_{\b(n)}~ \non \\
&\propto&\cD^\g \big (\D^{\b_1}_{[1]}{}_{\g}\D^{\b_2}_{[2]}{}_{\a_1} \cdots \D^{\b_n}_{[n]}{}_{\a_{n-1}}+ (n!-1)~\text{permutations} \big ) \F_{\b(n)}~. 
\eea
By making use of \eqref{Ident2}, it can be shown that the remaining $(n!-1)$ terms can be expressed in the same form as the first. Then transversality follows immediately as a consequence of property \eqref{TransOp}. However, $\mathbb{T}_{[n]}$ does not square to itself on $\mathds{V}_{(n)}$
\bea 
\mathbb{T}_{[n]} \mathbb{T}_{[n]} \F_{\a(n)} = \frac{1}{(2n+1)^n} \prod_{t=0}^{\lceil n/2 \rceil - 1}\big (\mathbb{F}+M^{(-)}_{(t,n)} \big )\prod_{t=1}^{\lfloor n/2 \rfloor }\big (\mathbb{F}-M^{(+)}_{(t,n)} \big )\mathbb{T}_{[n]} \F_{\a(n)}  ~,
\eea
where $M^{(\pm)}_{(t,n)}$ denotes the pseudo-masses associated with a partially massless superfield \eqref{PsudeoM}. We can immediately introduce the dimensionless operator  
\bea \label{superprojector}
\bm \P^{\perp}_{[n]}\F_{\a(n)} := (2n+1)^n \bigg [\prod_{t=0}^{\lceil n/2 \rceil - 1}\big (\mathbb{F}+M^{(-)}_{(t,n)} \big )\prod_{t=1}^{\lfloor n/2 \rfloor }\big (\mathbb{F}-M^{(+)}_{(t,n)} \big ) \bigg ]^{-1} \mathbb{T}_{[n]}\F_{\b(n)}~,
\eea
which is idempotent and transverse by construction. In addition, it can be shown that the operator $\bm \P^{\perp}_{[n]}$ acts as the identity on the space of transverse superfields \eqref{SIT}. Hence, $\bm \P^{\perp}_{[n]}$ satisfies properties \eqref{SProp} and can be identified as a rank-$n$ superprojector on AdS$^{3|2}$.

An alternative form of the superprojector $\bm \P^{\perp}_{[n]}$ can be derived, which instead makes contact with the Casimir operator $\mb{Q}$. Let us introduce the dimensionless operator 
\bea \label{altsupproj}
\widehat{\bm \P}{}^\perp_{[n]}\F_{\a(n)} &=& \bigg [\prod_{t=0}^{n-1}\big (\mathbb{Q}+\ri t \cS \cD^2 \big ) \bigg ]^{-1} \widehat{\D}^{\b_1}_{[1]}{}_{(\a_1}\widehat{\D}^{\b_2}_{[2]}{}_{\a_2} . . . \widehat{\D}^{\b_n}_{[n]}{}_{\a_n)}\F_{\b(n)}~,
\eea
where we denote  $\widehat{\D}^{\b}_{[j]}{}_{\a}$ as
\be \label{altop}
\widehat{\D}^{\b}_{[j]}{}_{\a}:= - \frac{\ri}{2}\cD^2 {\D}^{\b}_{[j]}{}_{\a}~.
\ee
 In the flat superspace limit, $\widehat{\bm \P}{}^\perp_{[n]}$ coincides with the superprojector derived in \cite{BHHK}. Making use of the properties of $\bm \P^{\perp}_{[n]}$ and the identity
\be \label{SCasF}
 - \frac{\ri}{2}\cD^2 \J_{\a(n)} = \frac{1}{2n+1} \big (\mathbb{F}  + 2n(n+2) \cS \big )\J_{\a(n)}~,
\ee
where $\J_{\a(n)}$ is an arbitrary transverse superfield, it can be shown that $\widehat{\bm \P}{}^\perp_{[n]}\F_{\a(n)}$ satisfies properties \eqref{SProp} and is also a superprojector on AdS$^{3|2}$. Using an analogous proof employed to show the coincidence of the two bosonic projectors in section \ref{secBFProj}, it can be shown that ${\bm \P}^\perp_{[n]}$ and $\widehat{\bm \P}{}^\perp_{[n]}$ are indeed equivalent. So far, we have been unable to obtain an expression for ${\bm \P}^\perp_{[n]}$ which is purely in terms of the Casmir operators $\mb{F}$ and $\mb{Q}$.

We recall that in the non-supersymmetric case, one starts with a field $\f_{\a(n)}$ lying on the mass-shell \eqref{OMS2} and its  projection $\P^{\perp}_{[n]}\f_{\a(n)}$ furnishes the reducible representation \eqref{red}. A single irreducible representation from the decomposition \eqref{red} can be singled out via application of the helicity projectors \eqref{helicityproj}. The significance of the condition \eqref{OMS2} is that it allows one to resolve the poles in both types of projectors.

 In the supersymmetric case, the equation analogous to \eqref{OMS2} which $\F_{\a(n)}$ should satisfy is \eqref{SOMS}. Upon application of $\bm \P^{\perp}_{[n]}$ on such a $\F_{\a(n)}$, one obtains the reducible representation \eqref{redS}. However, it appears that the imposition of \eqref{SOMS} does not allow one to resolve the poles of the superprojector in either of the forms \eqref{superprojector} or \eqref{altsupproj}. Therefore, rather then imposing \eqref{SOMS}, one must start with a superfield $\F_{\a(n)}$ obeying the first-order constraint \eqref{SFO}, which does allow for resolution of the poles. In this case, after application of $\bm \P^{\perp}_{[n]}$, the superfield $\F_{\a(n)}$ already corresponds to an irreducible representation with fixed superhelicity, relinquishing the need for superhelicity projectors. Thus, it suffices to provide only the superspin projection operators  $\bm \P^{\perp}_{[n]}$.

\subsection{Longitudinal projectors}
For $n\geq 1$, let us define the orthogonal compliment of $\bm \P^{\perp}_{[n]}$ acting on $\F_{\a(n)}$ by the rule
\be \label{SLongProj}
\bm \P^\parallel_{[n]}\F_{\a(n)} =  \big (\mathds{1}-\bm \P^{\perp}_{[n]} \big )\F_{\a(n)} ~.
\ee
By construction, the operators $\bm \P^{\perp}_{[n]}$ and $\bm \P^\parallel_{[n]}$ resolve the identity, $\mathds{1} = \bm \Pi^{\parallel}_{[n]} + \bm \Pi^{\perp}_{[n]}$, and are orthogonal projectors 
\be \label{SOrthoProjProp}
\bm \Pi^{\perp}_{[n]}\bm \Pi^{\perp}_{[n]} = \bm \Pi^{\perp}_{[n]}~, \qquad \bm \Pi^{\parallel}_{[n]}\bm \Pi^{\parallel}_{[n]} = \bm \Pi^{\parallel}_{[n]}~, \qquad \bm \Pi^{\parallel}_{[n]} \bm \Pi^{\perp}_{[n]} = \bm \Pi^{\perp}_{[n]} \bm \Pi^{\parallel}_{[n]}  = 0~. 
\ee
It can be shown that $\bm \P^\parallel_{[n]}$ extracts the longitudinal component of a superfield $\F_{\a(n)}$. A rank-$n$ superfield  $\J_{\a(n)}$ is said to be longitudinal if there exists a rank-$(n-1)$ superfield $\J_{\a(n-1)}$ such that $\J_{\a(n)}$ can be expressed as $\J_{\a(n)}= \ri^n\cD_\a \J_{\a(n-1)}$. Thus, we find
\be
\bm \P^\parallel_{[n]}\F_{\a(n)} = \ri^n \cD_{\a}\F_{\a(n-1)}~,
\ee
for some unconstrained real superfield $\F_{\a(n-1)}$. In order to see this, it proves beneficial to make use of the superprojector $\widehat{\bm \P}{}^{\perp}_{[n]}$, and express the operator $\widehat{\D}^{\b}_{[j]}{}_{\a}$ in the form
\be 
\widehat{\D}^{\b}_{[j]}{}_{\a}: = -\frac{1}{4}   \cD_\a \cD^\b \cD^2 + \big (\mathbb{Q}+\ri (j-1)\cS\cD^2\big )\d_\a{}^\b ~. 
\ee

 Using the fact that the $\bm \Pi^{\parallel}_{[n]}$ and $\bm \Pi^{\perp}_{[n]}$ resolve the identity, it follows that one can decompose any superfield $\F_{\a(n)}$ in the following manner
\be
\F_{\a(n)} = \F^{\perp}_{\a(n)} +\ri^n\cD_{\a}\F_{\a(n-1)}~.
\ee
Here, $\F^{\perp}_{\a(n)}$ is transverse and $\F_{\a(n-1)}$ is unconstrained. 
Repeating this prescription iteratively yields the decomposition
\bsubeq
\bea
\F_{\a(n)}&=& \sum_{j=0}^{\lfloor n/2 \rfloor} \big (  \cD_{\a(2)} \big )^j \F^{\perp}_{\a(n-2j)} + \ri^{n}\sum_{j=0}^{\lceil n/2 \rceil -1} \big ( \cD_{\a(2)} \big )^j \cD_\a \F^{\perp}_{\a(n-2j-1)}~.
\eea
\esubeq
Here, the real superfields $\F^{\perp}_{\a(n-2j)}$ and $\F^{\perp}_{\a(n-2j-1)}$  are transverse, except for $\F^{\perp}$.

It can be shown that the superprojector $\bm \P^{\perp}_{[n]}$ annihilates any longitudinal superfield. Indeed, let us consider the action of $\bm \P^{\perp}_{[n]}$ on a superfield $\J_{\a(n)}=\ri^n \cD_{\a}\L_{\a(n-1)}$. Opening the symmetrisation present in $\bm \P^{\perp}_{[n]}$ gives
\bea
\bm \P^{\perp}_{[n]}\J_{\a(n)} &=& \ri^n\D^{\b_1}_{[1]}{}_{(\a_1}\D^{\b_2}_{[2]}{}_{\a_2} . . . \D^{\b_n}_{[n]}{}_{\a_n)}\cD_{(\b_1} \L_{\b_2 ... \b_n)}~ \\
&=&\frac{\ri^n}{n!}\D^{\b_1}_{[n]}{}_{(\a_1}\D^{\b_2}_{[n-1]}{}_{\a_2} . . . \D^{\b_n}_{[1]}{}_{\a_n)}\big (\cD_{\b_n} \L_{\b_1 ... \b_{n-1}} + (n!-1)~\text{permutations} \big )~. \non
\eea 
Note that we have made use of the identity \eqref{Ident1} to rearrange the operators $\D^{\b}_{[j]}{}_{\a}$.
Making use of the relation \eqref{Ident3} allows us to express the other $(n!-1)$ permutations in the same form as the first. Then due to the property \eqref{TransOp}, it follows that 
\be \label{SPKL}
\J_{\a(n)}=\ri^n \cD_{\a}\L_{\a(n-1)}\qquad \implies \qquad \bm \P^{\perp}_{[n]}\J_{\a(n)} = 0~.
\ee
Consequently, the operator $\bm \P^\parallel_{[n]}$ acts as unity on the space of rank-$n$ longitudinal superfields $\J_{\a(n)}$
\be \label{LongIden}
\J_{\a(n)}=\ri^n \cD_{\a}\L_{\a(n-1)}\qquad \implies \qquad \bm \P^{\parallel}_{[n]}\J_{\a(n)} = \J_{\a(n)}~.
\ee


\subsection{Linearised higher-spin super-Cotton tensors}
In this section, we make use of the rank-$n$ superprojector to study the properties of superconformal higher-spin (SCHS) theories. In particular, we will make use of $\bm \P^{\perp}_{[n]}$ to construct the higher-spin super-Cotton tensors in AdS$^{3|2}$, which were recently derived in \cite{KP1}. The super-Cotton tensors $\mathfrak{W}_{\a(n)}(H)$ were shown to take the explicit form
\be \label{SCT}
\mathfrak{W}_{\a(n)}(H) =\D^{\b_1}_{[1]}{}_{(\a_1}\D^{\b_2}_{[2]}{}_{\a_2} \cdots \D^{\b_n}_{[n]}{}_{\a_n)}H_{\b(n)}~,
\ee
which is a real primary descendent of the SCHS superfield $H_{\a(n)}$. The latter is defined modulo gauge transformations of the form
\be \label{SCHSGT}
\d_{\L}H_{\a(n)} = \ri^n \cD_\a \L_{\a(n-1)}~,
\ee
where the gauge parameter $\L_{\a(n-1)}$ is a real unconstrained superfield.
 The super-Cotton tensor \eqref{SCT} satisfies the defining properties: (i) it is transverse 
\bsubeq \label{SCTP}
\be
\cD^\b\mathfrak{W}_{\b\a(n-1)}(H) = 0~; \\
\ee
and (ii) it is invariant under the gauge transformations \eqref{SCHSGT}
\be
 \mathfrak{W}_{\a(n)}(\d_{\L}H)= 0~.
\ee
\esubeq

The superprojectors \eqref{superprojector} can be used to recast the
 super-Cotton tensors \eqref{SCT} in the simple form
\be \label{SCTPr}
\mathfrak{W}_{\a(n)}(H) = \frac{1}{(2n+1)^n}\prod_{t=0}^{\lceil n/2 \rceil - 1}\big (\mathbb{F}+M^{(-)}_{(t,n)} \big )\prod_{t=1}^{\lfloor n/2 \rfloor }\big (\mathbb{F}-M^{(+)}_{(t,n)} \big )  \bm \P^{\perp}_{[n]}H_{\a(n)} ~,
\ee
where $M^{(\pm)}_{(t,n)}$ denotes the partial pseudo-masses \eqref{PsudeoM}. In the flat superspace limit, $\cS \rightarrow 0$,  the super-Cotton tensor \eqref{SCTPr} reduces to those given in \cite{K,KT}. Expressing $\mathfrak{W}_{\a(n)}(H) $ in the form \eqref{SCTPr} is beneficial for the following reasons: (i) transversality of $\mathfrak{W}_{\a(n)}(H)$ is manifest on account of property \eqref{STransprop}; (ii)  gauge invariance is also manifest as a consequence of \eqref{SPKL}; and (iii) in the transverse gauge
\be \label{TransG}
H_{\a(n)} \equiv H^{\text{T}}_{\a(n)}~, \qquad \cD^\b H^{\text{T}}_{\b\a(n-1)} = 0~,
\ee
it follows from \eqref{SIT} that $\mathfrak{W}_{\a(n)}(H)$ factorises as follows
\be \label{KinF}
\mathfrak{W}_{\a(n)}(H^\text{T}) = \frac{1}{(2n+1)^n}\prod_{t=0}^{\lceil n/2 \rceil - 1}\big (\mathbb{F}+M^{(-)}_{(t,n)} \big )\prod_{t=1}^{\lfloor n/2 \rfloor }\big (\mathbb{F}-M^{(+)}_{(t,n)} \big )  H^\text{T}_{\a(n)} ~.
\ee

From the above observations, it follows that the action \cite{KP2,KP3} for the superconformal higher-spin prepotential $H_{\a(n)}$
\begin{align}
\mb{S}^{(n)}_{\rm SCHS}[H] = 
- \frac{\ri^n}{2^{\left \lfloor{n/2}\right \rfloor +1}} \int \rd^3 x \rd^2 \theta  \, E \, H^{\a(n)} \mf{W}_{\a(n)}(H) ~, \qquad E^{-1}= {\rm Ber }(E_A{}^M)~,
\label{SCS}
\end{align}
is manifestly gauge-invariant. In the transverse gauge \eqref{TransG}, the kinetic operator in \eqref{SCS} factorises into wave operators associated with partially massless superfields of all depths, in accordance with \eqref{KinF}.


\section{Conclusion} \label{section4}

Given a maximally symmetric spacetime, the unitary irreducible representations of its isometry algebra may be realised on the space of tensor fields satisfying certain differential constraints. The purpose of a spin projection operator is to take an unconstrained field, which describes a multiplet of irreducible representations, and return the component corresponding to the irreducible representation with maximal spin.\footnote{In three dimensions, in order to single out an irreducible representation, one needs to bisect the spin projector into helicity projectors. } In this paper we have derived the spin projection operators for fields of arbitrary rank on AdS$_3$ space and their  extensions to $\mc{N}=1$ AdS superspace. We leave generalisations of our results to the $(p,q)$ AdS superspaces \cite{KLTM} with $\cN=p+q >1$  for future work. 

Making use of the (super)spin projection operators, we obtained new representations for the linearised higher-spin (super)Cotton tensors and the corresponding (super)conformal actions in AdS$_3$. The significance of these new realisations is that the following properties are each made manifest: (i) gauge invariance; (ii) transversality; and (iii) factorisation. We also show that the poles of the (super)projectors are intimately related to partially massless (super)fields. This property was first established in the case of AdS$_4$ (super)space in \cite{KP4, BHKP}, and appears to be a universal feature of the (super)projectors. It would be interesting to verify this in the case of AdS$_d$ with $d>4$. 

As compared with previous approaches in AdS$_4$ (super)space \cite{KP4, BHKP}, a novel feature of the spin projectors derived here is that they are formulated entirely in terms of Casimir operators of the AdS$_3$ algebra.\footnote{We were not able to obtain expressions for the superspin projection operators in AdS$^{3|2}$ which involve only Casimir operators.} Studying their zero curvature limit has allowed us to obtain new realisations of the spin projection operators in $3d$ Minkowski space in terms of only the Pauli-Lubanski scalar and the momentum squared operator. This idea may be straightforwardly applied to the case of 4$d$ Minkowski space to derive new realisations of the Behrends-Fronsdal projectors.   

In particular, let us define the square of the Pauli-Lubankski vector,
\begin{align}
\mathbb{W}^2=\mathbb{W}^a\mathbb{W}_a~,\qquad \mathbb{W}_a:=-\frac{1}{2}\ve_{abcd}M^{bc}\pa^d~. 
\end{align}
On the field $\phi_{\a(m)\ad(n)}$ of Lorentz type $(\frac{m}{2},\frac{n}{2})$, it may be shown that $\mathbb{W}^2$ assumes the form (see, e.g. \cite{Ideas})  
\begin{align}
\mathbb{W}^2\f_{\a(m)\ad(n)}=s(s+1)\Box\f_{\a(m)\ad(n)} +mn \pa_{\a\ad}\pa^{\b\bd}\f_{\a(m-1)\b\ad(n-1)\bd}~,
\end{align}  
where we have defined $s:=\frac{1}{2}(m+n)$. On any transverse field $\psi_{\a(m)\ad(n)}$ this reduces to $\big(\mathbb{W}^2-s(s+1)\Box\big)\psi_{\a(m)\ad(n)}=0$.
It is possible to express the Behrends-Fronsdal spin projection operators $\Pi^{\perp}_{(m,n)}$ solely in terms of the Casimir operators $\mathbb{W}^2$ and $\Box$ of the $4d$ Poincar\'e algebra as follows\footnote{These expressions may be easily converted to vector or four component notation.}
\begin{subequations} \label{4dProj}
\begin{align}
\Pi^{\perp}_{(m,n)}\f_{\a(m)\ad(n)}=\frac{m!}{(m+n)!n!}\frac{1}{\Box^n}&\prod_{j=0}^{n-1}\Big(\mathbb{W}^2-(s-j)(s-j-1)\Box\Big)\f_{\a(m)\ad(n)}\\
=\frac{n!}{(m+n)!m!}\frac{1}{\Box^m}&\prod_{j=0}^{m-1}\Big(\mathbb{W}^2-(s-j)(s-j-1)\Box\Big)\f_{\a(m)\ad(n)}~.
\end{align}
\end{subequations}
The operators $\Pi^{\perp}_{(m,n)}$ satisfy the four dimensional analogues of the properties \eqref{SPO}. 

In a similar fashion, it should be possible to obtain new realisations for the AdS$_4$ spin projection operators of \cite{KP4} in terms of the Casimir operators of the algebra $\mf{so}(3,2)$. In this case, $\Box$ should be replaced with the quadratic Casimir operator
\bea
\mb{Q}:=\Box_{\text{AdS}} - \mc{S}^2\big(M^2+\bar{M}^2\big)~, \qquad M^2:=M^{\a\b}M_{\a\b}~,
\quad  \bar{M}^2:=\bar{M}^{\ad\bd}\bar{M}_{\ad\bd}~. 
 \eea
 Finally, the role of $\mb{W}^2$ will be played by the quartic Casimir operator $\mb{W}^2_{\text{AdS}}$,\footnote{Here we use the convention $\big[\mc{D}_{\a\ad},\mc{D}_{\b\bd}\big]=-2\mc{S}^2\big(\ve_{\a\b}\bar{M}_{\ad\bd}+\ve_{\ad\bd}M_{\a\b}\big)$, where $\mc{S}^2$ is related to the AdS$_4$ scalar curvature via $R=-12\mc{S}^2$.}  
\bea
\mb{W}^2_{\text{AdS}}&:=&-\frac{1}{2}\big(\mb{Q}+2\mc{S}^2\big)\big(M^2+\bar{M}^2\big)+\mc{D}^{\a\ad}\mc{D}^{\b\bd}M_{\a\b}\bar{M}_{\ad\bd} \non \\
&&
-\frac{1}{4}\mc{S}^2\big(M^2M^2+\bar{M}^2\bar{M}^2+6M^2\bar{M}^2\big)~.
\eea
Both operators commute with the AdS$_4$ covariant derivative $\big[\mb{Q},\mc{D}_{\a\ad}\big]=\big[\mb{W}^2_{\text{AdS}},\mc{D}_{\a\ad}\big]=0$.\\

{\flushleft \textbf{Note added in proof:}}

When $m=n=s$, the spin projection operator \eqref{4dProj} takes the form
\begin{align}
\Pi^{\perp}_{(s,s)}\equiv \Pi^{\perp}_{(s)}=\frac{1}{\Box^s(2s)!}\prod_{j=0}^{s-1}\Big(\mb{W}^2-j(j+1)\Box\Big)~. \label{4dProjss}
\end{align}
In this case, it may be shown that $\Pi^{\perp}_{(s)}$ annihilates any field $\phi_{\a(s')\ad(s')}$ of lower rank:
\begin{align}
\Pi_{(s)}^{\perp}\phi_{\a(s')\ad(s')}=0~,\qquad s'<s~.\label{4dProjKill}
\end{align}
Let us comment on the implications of \eqref{4dProjKill} on fields with vectorial indices.
Consider a field $\bm h_{a_1\dots a_s}$ which is totally symmetric in its vector indices and has a non-zero trace
\begin{align}
\bm h_{a_1\dots a_s}= \bm h_{(a_1\dots a_s)}\equiv  \bm h_{a(s)}~,\qquad \eta^{bc}\bm h_{bca(s-2)}\neq 0~,
\end{align}
where $\eta_{ab}=\text{diag}(-1,1,1,1)$.
Upon converting to $4d$ two component spinor notation, see e.g. \cite{Ideas} for the details,
 $\bm h_{a(s)}$ decomposes into irreducible $\sSL(2,\mathbb{C})$ fields as follows
\begin{align}
\bm h_{\a_1\ad_1,\dots ,\a_s\ad_s}:= (\s^{a_1})_{\a_1\ad_1} \cdots (\s^{a_s})_{\a_s\ad_s}\bm h_{a_1\dots a_s} = h_{\a(s)\ad(s)}+\cdots~.
\end{align}
Here $h_{\a(s)\ad(s)}$ is associated with the traceless part of $\bm h_{a(s)}$, whilst the $+\cdots$ represent lower-rank fields $h_{\a(s')\ad(s')}$ associated with the trace of $\bm h_{a(s)}$. From \eqref{4dProjKill} it follows that the operator $\Pi^{\perp}_{(s)}$ selects the transverse and traceless (TT) component of $\bm h_{a(s)}$,
\begin{align}
\pa^b h^{\text{TT}}_{ba(s-1)}=0~,\qquad \eta^{bc}h^{\text{TT}}_{bca(s-2)}=0~,\qquad h^{\text{TT}}_{a(s)}:=\Pi^{\perp}_{(s)}\bm h_{a(s)}~.
\end{align}

Therefore, the spin-$s$ projection operator \eqref{4dProjss} is a TT projector when acting on a rank-$s$ field which is symmetric and traceful in its vectorial indices.
Similar conclusions hold in the three dimensional case.
 This is because the spin projection operators \eqref{SimpBosProj} and \eqref{SimpFermProj} in AdS$_3$ (and hence also those in $\mathbb{M}^3$ given by eq. \eqref{3dSPinproj}) satisfy a property analogous to \eqref{4dProjKill}, as pointed out in eqs. \eqref{BosProjProp} and \eqref{FermProjProp}.\\

\noindent
{\bf Acknowledgements:}\\
The work of DH is supported by the Jean Rogerson Postgraduate Scholarship and an Australian Government Research Training Program Scholarship. The work of SMK is supported in part by the Australian 
Research Council, project No. DP200101944. The work of MP is supported by the Hackett Postgraduate Scholarship UWA,
under the Australian Government Research Training Program.


\appendix

\section{Notation and conventions}\label{appendixA}

We follow the notation and conventions adopted in
\cite{KLTM1}. In particular, the Minkowski metric is
$\eta_{ab}=\mbox{diag}(-1,1,1)$.
The spinor indices are  raised and lowered using
the $\rm SL(2,{\mathbb R})$ invariant tensors
\bea
\ve_{\a\b}=\left(\begin{array}{cc}0~&-1\\1~&0\end{array}\right)~,\qquad
\ve^{\a\b}=\left(\begin{array}{cc}0~&1\\-1~&0\end{array}\right)~,\qquad
\ve^{\a\g}\ve_{\g\b}=\d^\a_\b
\eea
by the standard rule:
\bea
\psi^{\a}=\ve^{\a\b}\psi_\b~, \qquad \psi_{\a}=\ve_{\a\b}\psi^\b~.
\label{A2}
\eea

We make use of real gamma-matrices,  $\g_a := \big( (\g_a)_\a{}^\b \big)$, 
which obey the algebra
\be
\gamma_a \gamma_b=\eta_{ab}{\mathbbm 1} + \varepsilon_{abc}
\gamma^c~,
\label{A3}
\ee
where the Levi-Civita tensor is normalised as
$\varepsilon^{012}=-\varepsilon_{012}=1$. 
Given a three-vector $V_a$,
it  can be equivalently described by a symmetric second-rank spinor $V_{\a\b}$
defined as
\bea
V_{\a\b}:=(\g^a)_{\a\b}V_a=V_{\b\a}~,\qquad
V_a=-\hf(\g_a)^{\a\b}V_{\a\b}~.
\eea
Any
antisymmetric tensor $F_{ab}=-F_{ba}$ is Hodge-dual to a three-vector $F_a$, 
specifically
\bea
F_a=\hf\ve_{abc}F^{bc}~,\qquad
F_{ab}=-\ve_{abc}F^c~.
\label{hodge-1}
\eea
Then, the symmetric spinor $F_{\a\b} =F_{\b\a}$, which is associated with $F_a$, can 
equivalently be defined in terms of  $F_{ab}$: 
\bea
F_{\a\b}:=(\g^a)_{\a\b}F_a=\hf(\g^a)_{\a\b}\ve_{abc}F^{bc}
~.
\label{hodge-2}
\eea
These three algebraic objects, $F_a$, $F_{ab}$ and $F_{\a \b}$, 
are in one-to-one correspondence to each other, 
$F_a \leftrightarrow F_{ab} \leftrightarrow F_{\a\b}$.
The corresponding inner products are related to each other as follows:
\bea
-F^aG_a=
\hf F^{ab}G_{ab}=\hf F^{\a\b}G_{\a\b}
~.
\label{A.7}
\eea

The Lorentz generators with two vector indices ($M_{ab} =-M_{ba}$),  one vector index ($M_a$)
and two spinor indices ($M_{\a\b} =M_{\b\a}$) are related to each other by the rules:
$M_a=\hf \ve_{abc}M^{bc}$ and $M_{\a\b}=(\g^a)_{\a\b}M_a$.
These generators 
act on a vector $V_c$ 
and a spinor $\J_\g$ 
as follows:
\bea
M_{ab}V_c=2\eta_{c[a}V_{b]}~, ~~~~~~
M_{\a\b}\J_{\g}
=\ve_{\g(\a}\J_{\b)}~.
\label{generators}
\eea
The following identities hold:
\begin{subequations}
	\bea
	M_{\a_1}{}^{\b}\F_{\b \a_2 ... \a_n} &=& - \hf (n+2)\F_{\a(n)}~,\\
	M^{\b\g}M_{\b\g}  \F_{\a(n)} &=& -\hf n(n+2) \F_{\a(n)}
	~.
	\eea
\end{subequations}


\section{Generating function formalism}\label{appendixB}

We employ the generating function formalism which was developed in \cite{KP1}. Within this framework, a one-to-one correspondence between a homogenous polynomial $\f_{(n)}(\U)$ of degree $n$ and a rank-$n$ spinor field $\f_{\a(n)}$ is established via the rule
\be
\f_{(n)}(\U):=\U^{\a_1} \cdots\U^{\a_n}\f_{\a(n)}~.
\ee
Here, we have introduced the commuting real auxiliary variables $\U^{\a}$, which are inert under the action of the Lorentz generators $M_{\a\b}$. 

Making use of the auxiliary fields $\U^{\a}$, and their corresponding partial derivatives, $\partial_\b := \frac{\partial}{\partial \U^\b}$, we can realise the AdS$_3$ derivatives as index-free operators on the space of homogenous polynomials of degree $n$. We introduce the differential operators which increase and decrease the degree of homogeniety by $2$, $0$ and $-2$ respectively:
\be \label{Op}
\cD_{(2)} : = \U^\a \U^\b \cD_{\a\b}~, \quad \cD_{(0)}:= \U^\a \cD_\a{}^\b\partial_\b, \quad \cD_{(-2)}:= \cD^{\a\b}\pa_\a \pa_\b.
\ee
Note that the action of $\cD_{(0)}$ is equivalent to that of the Casimir operator $\cF$.

Making use of the algebra \eqref{ADSAlg}, one can derive the important identities
\begin{subequations}
	\begin{align}
	\big[\cD_{(2)}, \cD^{\phantom{.}t}_{(-2)}\big]\f_{(n)}&=4t(n-t+2)\big(\cQ-\tau_{(t,n+2)}\cS^2\big)\cD_{(-2)}^{t-1}\f_{(n)}~,\label{ID14}\\
	\big[\cD_{(-2)}, \cD^{\phantom{.}t}_{(2)}\big]\f_{(n)}&=-4t(n+t)\big(\cQ-\tau_{(t,n+2t)}\cS^2\big)\cD_{(2)}^{t-1}\f_{(n)}~,\label{ID15}\\
\cD^t_{(2)}\cD^t_{(-2)}\f_{(n)} &= \prod_{j=0}^{t-1}\Big (\cF^2 - \big(n-2j \big )^2 \big (\cQ-(n-2j-2) (n-2j+2)\cS^2 \big )  \Big ) \f_{(n)}~, \label{PTI}
\end{align} 
\end{subequations}
via induction on $t$. Here $\cQ$ and $\cF$ are the quadratic Casimir operators \eqref{QC} and $\t_{(t,n)}$ are the partially massless values \eqref{PMV}.


\begin{footnotesize}
	
\end{footnotesize}
\end{document}